\documentclass[twocolumn,twocolappendix]{aastex631}

\usepackage{amsmath}
\usepackage{empheq}
\usepackage{graphicx}
\expandafter\let\csname equation*\endcsname\relax
  \expandafter\let\csname endequation*\endcsname\relax 
\usepackage{mathrsfs}
\usepackage{textcomp}
\usepackage{enumitem}   
\usepackage{gensymb}
\usepackage{hyperref}
\usepackage{graphicx}
\usepackage[caption=false]{subfig}
\usepackage{placeins}
\makeatletter
\AtBeginDocument{%
  \setlength{\textfloatsep}{8pt}%
  \setlength{\dbltextfloatsep}{8pt}%
  \setlength{\floatsep}{8pt}%
  \setlength{\dblfloatsep}{8pt}%
  \setlength{\intextsep}{8pt}%
  \setlength{\abovedeluxetableskip}{0pt}%
  \setlength{\belowdeluxetableskip}{0pt}%
}
\makeatother
\usepackage{multirow}
\usepackage{longtable}
\usepackage[flushleft]{threeparttable}
\usepackage{booktabs} 
\usepackage{CJK}
\usepackage{verbatim}

\hypersetup{colorlinks, linkcolor={blue}, citecolor={blue}, urlcolor={blue}}

\definecolor{DarkOrange}{RGB}{204, 85, 0}
\definecolor{LincolnGreen}{RGB}{17, 102, 0}

\usepackage{xspace}

\newcommand\chandra{\textit{Chandra}\xspace}

\newcommand\hst{\textit{HST}\xspace}

\newcommand\W {{W^r_{\ \phi}}}

\newcommand\txmm{3XMM J215-05\xspace}

\begin{document}
\pagenumbering{arabic}

\title{Demographics of Off-Nuclear Tidal Disruption Events Powered by Wandering Black Holes  }
\author[0000-0002-5063-0751]{Muryel Guolo}
\affiliation{School of Natural Sciences, Institute for Advanced Study, 1 Einstein Drive, Princeton, NJ 08540, USA}
\affiliation{Bloomberg Center for Physics and Astronomy, Johns Hopkins University, 3400 N. Charles St., Baltimore, MD 21218, USA}

\begin{abstract}

The four recently discovered off-nuclear tidal disruption events (TDEs)---EP240222a, AT2024tvd, AT2025abcr, and AT2023mfm---together with those found serendipitously over past decades (ESO 243-49 HLX-1 and 3XMM J2150-05), reveal a small but robust population of accreting off-nuclear, or ``wandering'' black holes (WBHs). We analyze their demographic properties in comparison to other populations, we find that (i) every event occurs in a massive early-type galaxy within the narrow range $10.7 \lesssim \log_{10}(M_\star/M_\odot) \lesssim 11.1$, a range that very rarely hosts nuclear TDEs; (ii) their radial displacements span $1 \lesssim R_{\bullet}/{\rm kpc} \lesssim 40$, or $0.003 \lesssim R_{\bullet}/R_{200} \lesssim 0.1$; (iii) a stellar counterpart is detected in pre-TDE observations only when located outside the galaxy's half-light radius; (iv) while nuclear TDE hosts are usually isolated, off-nuclear TDEs and quiescent WBHs are more likely to inhabit galaxy groups or clusters. Using the \texttt{ROMULUS} cosmological simulation, we show that all of these follow naturally from hierarchical galaxy formation. We construct the volumetric density of WBHs, $\phi_{\rm WBH}(M_{\star})$, which places the central 68\% of local WBHs in galaxies with $10.7 \lesssim \log_{10}(M_\star/M_\odot) \lesssim 11.3$, explaining (i) and (iv). This WBH distribution can imply a higher off-nuclear than nuclear TDE rate, at these galaxy masses. Their intrinsic radial distribution is Gaussian-like in $\log_{10}(R_\bullet/R_{200})$, consistent with (ii) except for a deficit of sources at the largest separations. Combined with parent-galaxy light, which suppresses counterpart detections at small separations, this distribution also explains (iii). These results show that off-nuclear TDEs are the accreting counterparts of the WBHs long predicted by cosmological simulations, and that they trace large-scale structure.
\end{abstract}


\vspace{1em}

\section{Introduction}\label{sec:intro}
It is now established that nearly all massive galaxies host a massive black hole (MBH) at their centers \citep{Kormendy2013}. MBH masses ($M_{\bullet}$) correlate strongly with the properties of their host galaxies' central regions \citep{Magorrian1998,Ferrarese2000,Tremaine2002,Gultekin2009}, indicating a co-evolution of black holes and galaxies over cosmic time \citep[e.g.,][]{Volonteri2003,Natarajan2014}.

In the standard paradigm of galaxy formation, MBHs\footnote{Here MBHs are defined as black holes not formed through standard stellar evolution, i.e., with masses $M_{\bullet} \gg 10^{3}\,M_\odot$, including intermediate-mass black holes (IMBHs; $10^{3} \ll M_{\bullet} \lesssim 10^{5}\,M_\odot$) and supermassive black holes (SMBHs; $M_{\bullet} \gg 10^{5}\,M_\odot$).} grow primarily through gas accretion in galactic nuclei, producing active galactic nuclei (AGN), while galaxies assemble hierarchically through mergers. Following a merger, the satellite galaxy and its central MBH are left on a wide orbit in the remnant and must sink from galactic scales toward the center of the primary.

That decay involves multiple physical processes operating across wide spatial and temporal scales \citep{Begelman1980,Colpi2014}. When the sinking is efficient, the two MBHs can form a bound binary and eventually coalesce, producing gravitational waves detectable by space-based interferometers such as the Laser Interferometer Space Antenna (LISA) \citep{AmaroSeoane2017}. In many systems, however, orbital decay stalls at kiloparsec scales, where dynamical friction dominates \citep{Chandrasekhar1943} and the sinking time can exceed a Hubble time for sufficiently low-mass MBHs. As a result, a substantial population of off-nuclear MBHs is expected to persist to low redshift.

\begin{figure*} \centering \includegraphics[width=\linewidth]{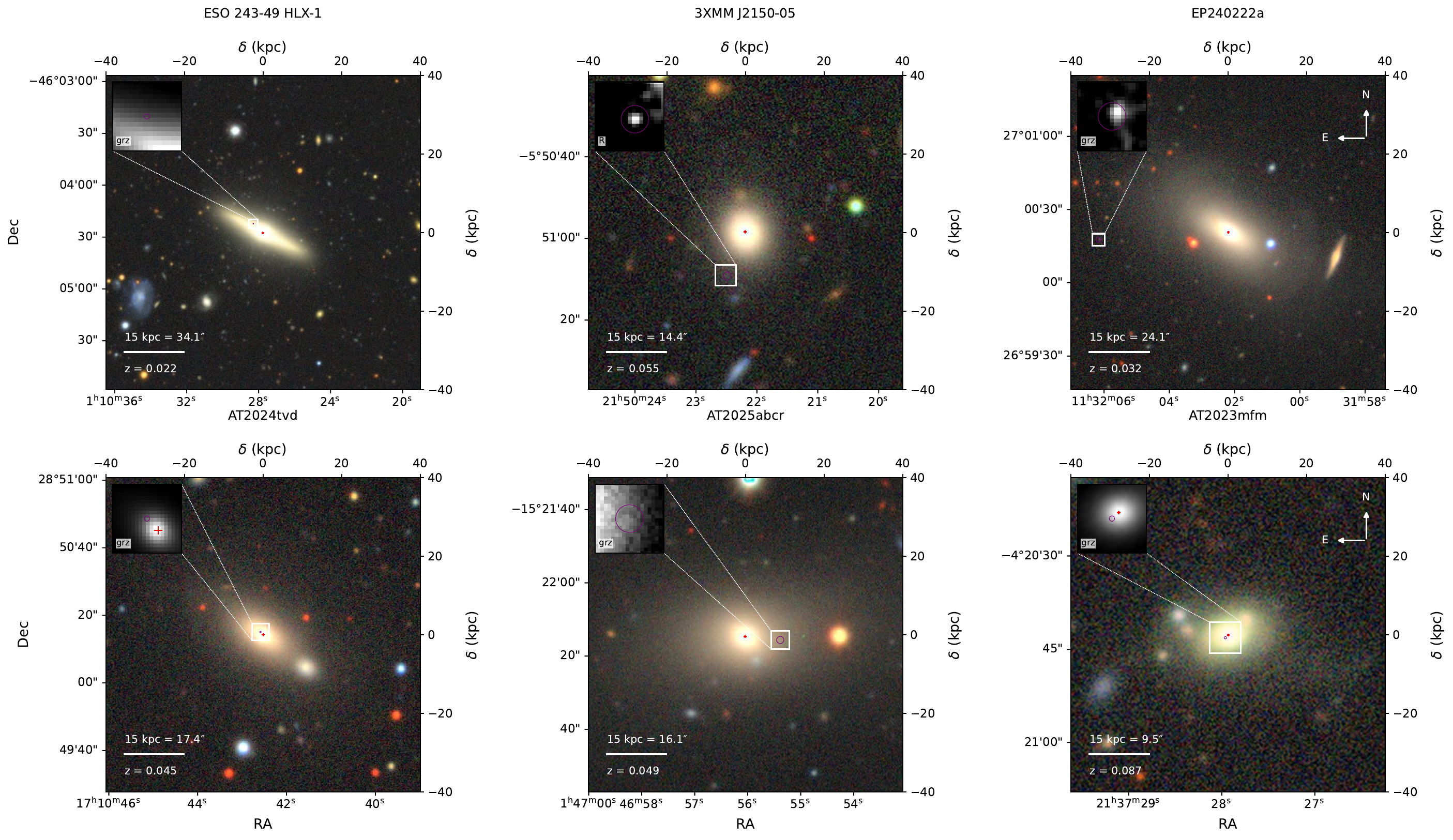} 

\caption{Optical ground-based images of the parent galaxies listed in Table~\ref{tab:hosts}. Each panel shows an $80\times80$~kpc field centered on the galaxy nucleus, with a zoomed-in view of the transient position. Purple circles mark the most precise positions, derived from space-based X-ray or ultraviolet observations. Pronounced stellar overdensities are detected for 3XMM~J2150$-$05 and EP240222a, whereas no resolved stellar counterpart is visible at the locations of the remaining sources above the smooth halo light.} \label{fig:1} \end{figure*}

The existence of these long-lived, off-nuclear MBHs was predicted by semi-analytical calculations, well before modern hydrodynamical simulations, as a natural consequence of hierarchical galaxy assembly and inefficient orbital decay \citep[e.g.,][]{Governato1994,Volonteri2003,Islam2004,Bellovary2010}.

Cosmological simulations are now the primary theoretical tool for studying this population of wandering black holes (WBHs), here defined as MBHs that are not located at the center of their main galactic halo, or that reside in satellite substructures. In particular, this requires simulations that seed black holes based on local gas properties, incorporate subgrid prescriptions for dynamical friction, and allow MBHs to orbit freely rather than being artificially repositioned at halo centers. Notable examples include the \texttt{ROMULUS} \citep{Tremmel2017} and \texttt{ASTRID} \citep{Ni2022} simulations.

Studies based on \texttt{ROMULUS} find a roughly linear scaling between the number of WBHs per halo and the halo virial mass, with Milky Way–like halos hosting of order ten WBHs \citep{Tremmel2018wanderings} and massive galaxy clusters hosting more than a thousand \citep{Ricarte2021a}. Despite their predicted abundance, WBHs are expected to accrete weakly in the local Universe, with typical Eddington ratios $\lesssim10^{-4}$ \citep{Ricarte2021b}, owing to gas-poor environments and the lack of efficient angular-momentum diffusion mechanisms, such as those available to secularly fed central SMBHs \citep{Storchi-Bergmann2019}. The same weak accretion is found for WBHs in \texttt{ASTRID} \citep{Weller2026}, and in \texttt{ROMULUS} it follows from the time spent off-center \citep{Whitaker2026}. This makes them extremely difficult to detect directly, and observationally WBHs have therefore proven elusive.

\begin{deluxetable*}{lcc|ccccc|cc}
\tablecaption{Parent galaxies and halo properties of off-nuclear TDEs \label{tab:hosts}}
\tabletypesize{\scriptsize}
\tablehead{
\multicolumn{3}{c|}{Source} & \multicolumn{5}{c|}{Parent Galaxy} & \multicolumn{2}{c}{Stellar Counterpart} \\
\colhead{Name} & \colhead{$R_{\bullet,{\rm p}}$} & \multicolumn{1}{c|}{$R_{\bullet}$} &
\colhead{$z$} & \colhead{$\log_{10} M_\star$} & \colhead{$R_{1/2}$} & \colhead{$\log_{10} M_{200}$} & \multicolumn{1}{c|}{$R_{200}$} &
\colhead{$M_{\star,{\rm scp}}$} & \colhead{Detected?} \\
\colhead{} & \colhead{(arcsec, kpc)} & \multicolumn{1}{c|}{(kpc)} &
\colhead{} & \colhead{($M_\odot$)} & \colhead{(arcsec, kpc)} & \colhead{($M_\odot$)} & \multicolumn{1}{c|}{(kpc)} &
\colhead{($M_\odot$)} & \colhead{ground$|$space}
}
\startdata
ESO 243--49 HLX-1 & 8.0, 3.7 & $4.2_{-0.5}^{+2.5}$ & 0.022 & $10.73\pm0.10$ & 8.9, 4.1 & $12.4\pm0.2$ & $270_{-40}^{+50}$ & $10^{5.5}$--$10^{6.5}$ & $\times/\checkmark$ \\
3XMM J2150-05 & 11.6, 12.6 & $14.5_{-1.8}^{+8.5}$ & 0.055 & $10.83\pm0.10$ & 2.3, 2.5 & $12.5\pm0.2$ & $300_{-40}^{+60}$ & $10^{7.3\pm 0.3}$ & $\checkmark/\checkmark$ \\
EP240222a & 53.1, 34.4 & $39.8_{-4.9}^{+23.4}$ & 0.032 & $10.95\pm0.10$ & 12.1, 7.8 & $12.9\pm0.2$ & $410_{-70}^{+80}$ & $10^{6.8\pm 0.1}$ & $\checkmark/\checkmark$ \\
AT2024tvd & 0.92, 0.83 & $0.95_{-0.12}^{+0.56}$ & 0.045 & $10.91\pm0.11$ & 5.3, 4.8 & $12.8\pm0.2$ & $370_{-60}^{+70}$ & $<10^{7.6}$ & $\times/\times$ \\
AT2025abcr & 9.5, 9.2 & $10.7_{-1.3}^{+6.3}$ & 0.049 & $11.08\pm0.10$ & 8.0, 7.8 & $13.2\pm0.2$ & $510_{-80}^{+100}$ & $<10^{7.4}$ & $\times/\times$ \\
AT2023mfm & 0.66, 1.1 & $1.26_{-0.15}^{+0.74}$ & 0.087 & $10.80\pm0.10$ & 3.4, 5.6 & $12.5\pm0.2$ & $300_{-40}^{+60}$ & $10^{7}$--$10^{8}$ & $\times/\checkmark$ \\
\enddata
\end{deluxetable*}

Non-accreting WBHs have been inferred dynamically in the remnant nuclei of tidally stripped satellites. At the high-mass end, these nuclei are observed as ultra-compact dwarfs (UCDs), which are common in galaxy clusters \citep{Pfeffer2014,Voggel2019}. UCDs, and many of the most massive star clusters in halos from the Milky Way to galaxy clusters, are thought to be the tidally stripped nuclear star clusters of disrupted dwarfs \citep{Pfeffer2016,Mayes2021,Mayes2024,Neumayer2020,Pfeffer2021}. Black holes have been inferred not only in the most massive cluster UCDs \citep{Seth2014,Ahn2017,Ahn2018,Afanasiev2018,Taylor2025}, but also in stripped nuclei in M31 \citep{Gebhardt2005,Pechetti2022} and in $\omega$~Centauri, the most massive stellar cluster in the Milky Way \citep{Haberle2024}.

Accreting WBHs (of intermediate mass)
have been invoked as engines of some ultra-luminous X-ray sources (ULXs;
$10^{39} \lesssim L_{\rm X} \lesssim 10^{41}\,\mathrm{erg\,s^{-1}}$), although  objects in this
class are now understood to be extreme stellar-mass (mostly neutron stars) accretors
\citep{Kaaret2017}.
These have also been proposed to appear as off-nuclear AGN, particularly in merging systems \citep{Comerford2009,Comerford2014,Reines2020,Ward2021,Uppal2024}, but such sources are often ambiguous \citep[e.g.,][]{Sturm2026}, being difficult to distinguish from recoiling SMBHs or from dual, similar-mass AGN produced by major mergers. A secure AGN detection is lacking in galaxies with $M_\star \lesssim 10^{8}\,M_\odot$, where the published candidates are identified from emission-line ratios, infrared colors, or X-ray luminosities that stellar-mass accretors can also produce \citep{Marleau2017,Mezcua2018,FerreMateu2021}.

A previously quiescent WBH can be detected, however, in the unlikely event that a star is disrupted by it. Such tidal disruption events \citep[TDEs,][]{Rees1988,Gezari2021} can supply the accreting material needed to briefly power luminous emission, well beyond what any stellar-mass accretor can produce. Such an off-nuclear TDE, however, can happen only if the WBH retains a bound compact stellar cluster. The reason for that is that the TDE rate is a strong function of the stellar density within the MBH's radius of influence. An MBH with no bound compact stellar counterpart, only interacting with halo, bulge, or disk stars, has a negligible TDE rate \citep{Stone2012,Pfister2020,Chang2025b}. Fortunately, even during the tidal stripping in a minor merger, an accreted galaxy should retain its nuclear stellar cluster, because its density is usually high enough to survive the tidal forces that strip the outer envelope \citep{Bekki2001,Bekki2003,Pfeffer2013,Van_Wassenhove2014,Tremmel2018binary}.

Off-nuclear TDEs therefore provide one of the most direct
and unambiguous probes of WBHs, particularly at distances beyond the extreme Local Universe ($D \gg 20$~Mpc), where dynamical techniques are unfeasible. Unlike for AGN, $M_{\bullet}$ in TDEs can be inferred directly from multi-wavelength modeling based on first-principles models, independent of any scaling relations (such as those based on single-epoch broad-line emission measurements, which are poorly calibrated at the low-mass end of interest here), enabling, e.g., black hole--host scaling relations to be recovered independently \citep[e.g.,][]{Mummery2024,Guolo2025d}.

The first clear case of an accreting WBH is ESO 243--49 HLX-1 \citep{Farrell2009,Webb2014,Soria2017,Lasota2011,Soria2026,Yu2026}. Its transient, but recurrent, light curve may not be immediately identified with the simple one-off transient associated with an off-nuclear TDE, so the origin of its emission has been disputed between a persistent but unstable accretion disk \citep[e.g.,][]{Soria2017,Soria2026} and the repeated stripping of a star \citep[i.e., a repeating partial TDE;][]{Lasota2011,Yu2026}.

The discovery of \txmm\ in 2018 \citep{Lin2018,Lin2020} marked perhaps the first undisputed off-nuclear TDE. Over the past two years, four additional sources have been identified: EP240222a \citep{Jin2025} by \textit{Einstein Probe} \citep{Yuan2018}, and AT2024tvd \citep{Yao2025}, AT2025abcr \citep{Stein2026}, and AT2023mfm \citep{Sfaradi2026,Li2026} by the Zwicky Transient Facility \citep[ZTF;][]{Bellm2019}, with a few other candidates depending on the selection criteria and data-quality requirements one imposes \citep[e.g.,][]{Donato2014,Grotova2025,Chang2025}.

Despite the small current sample, some interesting demographic properties are already apparent. All events occur in massive, early-type parent galaxies with total stellar masses ($M_{\star}$) of order $10^{11}\,M_\odot$. They show a wide range of projected distances from their parent-galaxy nuclei, $0.8 \lesssim R_{\bullet,{\rm p}}/{\rm kpc} \lesssim 35$. In addition, in pre-TDE ground-based images, the sources at larger distances have a detected stellar counterpart (either a UCD or a massive stellar cluster) with stellar counterpart masses ($M_{\star,\mathrm{scp}}$) $\sim 10^{7}\,M_{\odot}$, while those occurring closer to the halo center lack a detectable stellar counterpart. In some cases, even space-based observations still do not detect a stellar counterpart.

In this \textit{Letter}, we present a study of the demographic properties of off-nuclear TDEs, first in comparison with other relevant MBH populations (nuclear TDEs and dynamically detected WBHs), and then show, using the \texttt{ROMULUS} simulation, that their observed demographic properties arise naturally from hierarchical galaxy and black hole assembly.

In \S\ref{sec:data} we describe the off-nuclear TDE sample and its properties, the comparison samples, and the \texttt{ROMULUS} simulation; in \S\ref{sec:results_obs} we present the observational comparison between the samples; in \S\ref{sec:results_sim} we compare the simulation with the observations,
and we conclude in
\S\ref{sec:conclusion}. In this paper we adopt a flat $\Lambda$CDM cosmology, with $H_0 = 69$ km s$^{-1}$ Mpc$^{-1}$ \citep{planck}; uncertainties (error bars and contours) are reported as 68\% credible intervals, unless otherwise stated.

\section{Samples, Observed Properties, and Simulation}\label{sec:data}

\subsection{Off-Nuclear TDEs}\label{sec:sample}
Our sample consists of six sources: ESO 243--49 HLX-1, \txmm, EP240222a, AT2024tvd, AT2025abcr, and AT2023mfm.
These sources share several defining properties:

\begin{itemize}

    \item transient multi-wavelength emission, with detections spanning X-rays (except for AT2023mfm) through the UV and optical;
    \item peak luminosities well above $10^{42}\,\mathrm{erg\,s^{-1}}$, including in the X-rays for every detected source, which excludes stellar-mass accretors and supernovae;
    \item for the X-ray-detected sources, a super-soft, thermal X-ray spectrum;
    \item for the sources discovered in real time, optical spectra showing a sustained, extremely blue continuum, transient broad Balmer emission lines and, in some cases, higher-ionization features such as He~II and Bowen fluorescence, which also exclude supernovae.
\end{itemize}

\noindent Crucially, however, all six sources are significantly displaced from the
centers of their parent galaxies\footnote{The word ``host'' is usually used for the galaxy in which a transient, including a TDE, originates. For an off-nuclear TDE that wording can be ambiguous. We use here ``parent galaxy'' for the massive galaxy, and ``stellar counterpart'' for the dwarf galaxy or massive stellar cluster on which the WBH is centered.}. Individually, these properties are characteristic of
TDEs \citep[e.g.,][]{van_Velzen_21,Charalampopoulos2022,Yao2023,Guolo2024,Mummery2024,Grotova2025,Mummery_vanVelzen2024}
and, taken together, make these events the most secure cases of
off-nuclear TDEs currently known. The addition of ESO 243--49 HLX-1 to the sample may be the most contentious. Here we rely on the arguments of \citet{Lasota2011} that the alternative of persistent accretion through an unstable disk is not favored over repeated tidal stripping of a star, but primarily on the work of \citet{Yu2026}, which shows that ESO 243--49 HLX-1's light curve and evolving spectral energy distribution can be reproduced by repeated realizations (i.e., a repeating partial TDE) of the same time-dependent accretion-disk model solution which has been very successful in explaining observations of TDEs at the population level \citep[e.g.,][]{Mummery2020, Mummery2024, Goodwin2026, Guolo2025d}.

\begin{figure}[tbp]
    \centering
    \includegraphics[width=\linewidth]{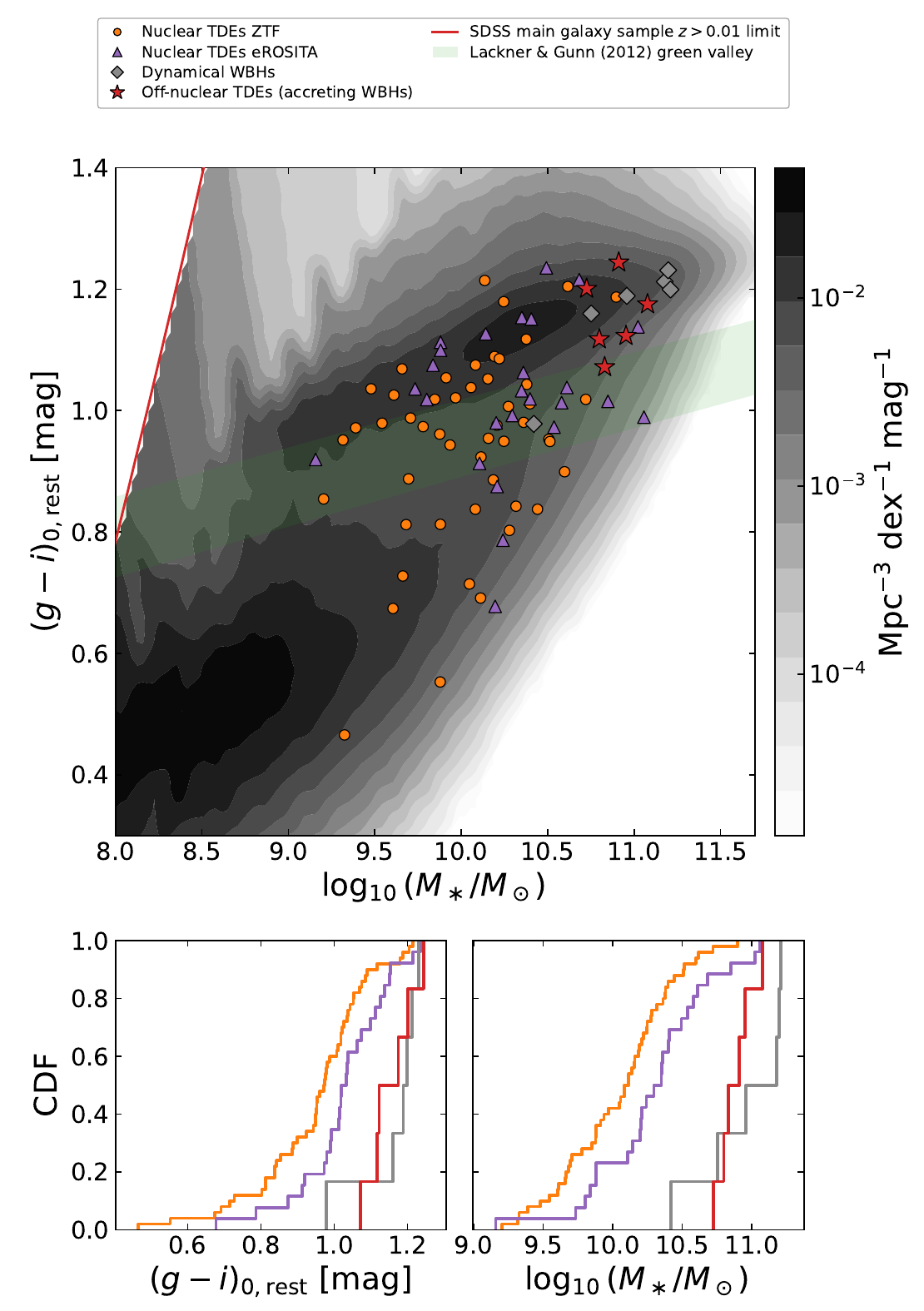}
  \caption{Rest-frame intrinsic $(g-i)$ color and $M_{\star}$. Orange circles are nuclear TDEs from ZTF, and purple triangles are nuclear TDEs from eROSITA. Red stars are the parent galaxies of the off-nuclear TDEs, and gray diamonds are the parents of the dynamically detected WBHs. The gray background is the volume density of SDSS galaxies. The dashed curve is the SDSS main-galaxy-sample limit at $z>0.01$, and the green band marks the green valley \citep{Lackner2012}. The lower panels are the cumulative distributions of color and $M_{\star}$.}
  \label{fig:mass-color}
\end{figure}

Two additional sources, eRASS~J1421$-$29 \citep{Grotova2025} and 4XMM~J1615$+$19
(also known as NGC~6099~HLX-1; \citealt{Chang2025}), are, in our view, possibly
genuine off-nuclear TDEs. However, the published analyses and the
currently available data are not yet sufficient to conclusively rule out alternative
explanations. We therefore do not include them in our sample\footnote{Importantly,
both sources seem to be associated with massive early-type galaxies with stellar
masses similar to those of our main sample. Including them would not alter our
conclusions regarding the galaxy-mass preference discussed in
\S\ref{sec:parent_pref}.}. A third source, WINGS~J1348$+$26 \citep{Maksym2013,Donato2014}, is a TDE originating
from the nucleus of a compact galaxy ($M_{\star}\sim10^{8}\,M_{\odot}$),
spectroscopically confirmed to be associated with the galaxy cluster
Abell~1795 \citep{Maksym2014}. In some sense, this event could be considered
``off-nuclear'' with respect to the cluster center and its brightest cluster galaxy (BCG). At this point,
however, this classification becomes largely a matter of semantics and of where one draws an arbitrary boundary\footnote{For the sake of argument, consider a TDE associated with the nucleus of a galaxy with $M_\star \sim 10^{10}\,M_\odot$ in the halo of a BCG with $M_\star > 10^{11}\,M_\odot$. Calling that an off-nuclear TDE would be highly debatable, so there must be an arbitrary cut between nuclear and off-nuclear TDEs.}. We choose not to include this source in our sample.

We also do not include luminous fast blue optical transients
\citep[LFBOTs; e.g.,][]{Perley2019,Ho2023,Somalwar2025}.
Although these sources have been interpreted as TDEs by some authors, their
observed properties—including light-curve evolution and spectral characteristics—are fundamentally distinct from those of the
off-nuclear TDEs analyzed here, suggesting a different physical origin (see also \S\ref{sec:parent_pref} for the differences in their host galaxies).

Figure~\ref{fig:1} shows pre-TDE color
images of the six parent galaxies, with zoomed-in regions indicating the TDE
locations. Images are from the Legacy Surveys \citep{Dey2019}, taken either pre-transient or more than a decade after the transient. The purple circles mark the most precise transient positions, derived
from space-based UV or X-ray observations using either \chandra\ ($\sim1\arcsec$ point-spread function, PSF) or \hst\ ($\sim0.1\arcsec$ PSF). From these ground-based images, clear stellar overdensities are visible only at the locations of \txmm\ and EP240222a, and not for the other four sources. 

In Table~\ref{tab:hosts}, we summarize the basic parent-galaxy properties: redshift
($z$); projected ($R_{\bullet,{\rm p}}$) and deprojected ($R_{\bullet}$) offsets of the
TDEs from their parent nuclei; $M_{\star}$ and virial ($M_{200}$) masses;
half-light ($R_{1/2}$) and virial ($R_{200}$) radii. Projected offsets are taken
from the discovery papers, while deprojected distances assume an isotropic
orientation distribution, i.e., $R_{\bullet} = \frac{R_{\bullet,{\rm p}}}{\rm sin (\theta)}$, with  $P(\theta)\propto\cos\theta$. Half-light radii are taken
from the Legacy Surveys Data Release~10 \citep{Dey2019}, and virial masses and radii are inferred
from $M_{\star}$ and $R_{1/2}$ as described in Appendix~\ref{app:1}.

Rather than relying on heterogeneous $M_{\star}$ estimates from the literature, which use distinct stellar-population models and apertures, we measure parent-galaxy $M_{\star}$ in a uniform manner. As described in Appendix~\ref{app:2}, we adopt a color-dependent mass-to-light ratio ($M_{\star}/L_{i}$) applied to S\'ersic-profile luminosities, following the same approach used by \citet{Baldry2012} and \citet{Taylor2011} to derive the local galaxy stellar mass function (GSMF). We include both a systematic uncertainty of 0.1~dex, associated with the different $M_{\star}$ that can produce the same color \citep{Taylor2011}, and the uncertainty in the photometry; the former dominates the latter for massive galaxies. This choice is essential for our comparison with the GSMF predictions in \S\ref{sec:results_sim}, and it allows rapid mass measurements for the large samples in \S\ref{sec:comp_samp}.

Table~\ref{tab:hosts} also lists the stellar counterparts of the WBHs. We first ask whether a counterpart is detected in pre-TDE, ground-based imaging, at the depth and seeing of a deep all-sky survey. Only EP240222a and 3XMM~J2150$-$05 meet this criterion, and we measure their masses with the same method as the parent galaxies (Appendix~\ref{app:2}), quoting a central value and its uncertainty. We then ask whether post-transient space-based imaging (\hst\ or \textit{JWST}) requires a stellar component in addition to the transient light. We judge this to be the case for ESO~243--49 HLX-1 in the 2022 \hst\ data \citep[][Yu et al., in prep.]{Soria2026} and for AT2023mfm in the 2026 \hst\ data \citep{Sfaradi2026}. These masses are less straightforward. For HLX-1 they depend on the modeling of both the accretion emission and the stars, and the systematic uncertainty dominates: different choices give statistically distinct values, all contained in the range $10^{5.5}$--$10^{6.5}\,M_{\odot}$ that we adopt \citep[][Yu et al., in prep.]{Soria2026}. For AT2023mfm, stellar light is required only for the F625W photometry of \citet{Sfaradi2026}; the UV bands are consistent with disk emission, so this observed single-band flux allows for masses in the $10^{7}$--$10^{8}\,M_{\odot}$ range. For AT2024tvd and AT2025abcr, the \hst\ and \textit{JWST} images were taken only months after discovery. The entire SED can still be explained by the transient, and stars do not dominate the emission at any wavelength where the point source is detected\footnote{A stellar counterpart to AT2024tvd was claimed by \citet{Patra2025}. \citet{Guolo2026} shows, however, that this is an artifact of their fitting method.}. We therefore use the upper limits from the discovery papers, $\lesssim 10^{7.6}\,M_{\odot}$ and $\lesssim 10^{7.4}\,M_{\odot}$, respectively.

We do not present $M_{\bullet}$ estimates for the full sample. A reliable and
uniform method capable of recovering black hole–host scaling relations has not
yet been applied to all four sources. The method of
\citet{Guolo_Mummery2025} has been applied to AT2024tvd and \txmm\
\citep{Guolo2025d,Guolo2026}, finding that the black hole powering
AT2024tvd is $\sim10^{6}/10^{4.4}\simeq40$ times more massive than
that of \txmm. Comparable estimates are unavailable for the remaining events,
precluding a homogeneous comparison. Moreover, in the \texttt{ROMULUS}
simulations all black holes are seeded with the same initial mass,
$M_{\bullet}=10^{6}\,M_{\odot}$, preventing a direct comparison between simulated
and observed $M_{\bullet}$.

\subsection{Comparison Samples}\label{sec:comp_samp}
Here we construct comparison samples of MBHs against which to compare the off-nuclear TDE sample just presented. The most direct comparison is with nuclear TDEs. Because the off-nuclear sample contains both X-ray- and optically selected sources, and in order to keep the selection function of the nuclear sample reasonably uniform, we draw it from the surveys that have discovered the most TDEs in the optical and in X-rays, ZTF \citep{Bellm2019} and the Russian--German eROSITA survey \citep[eROSITA;][]{Predehl2021}, respectively.

For ZTF we start from the TDE sample within the ZTF Bright Transient Survey \citep[BTS;][]{Perley2020}, a flux-limited, spectroscopically complete survey within ZTF. We exclude LFBOTs \citep{Perley2019,Ho2023,Somalwar2025} that were mistakenly flagged as TDEs in BTS, and we limit the sample to $z \leq 0.2$. This keeps the comparison in the local Universe, not much beyond the most distant off-nuclear TDE, while still retaining a non-negligible number of the rare nuclear TDEs in massive galaxies, which are discovered at higher redshift on average. This leaves 50 nuclear TDEs from ZTF. For eROSITA we start from the published samples of the Russian and German portions of the survey, \citet{Sazonov2021} and \citet{Grotova2025}, respectively. We restrict the sample to sources whose hosts have a spectroscopic redshift, and further to $z \leq 0.2$. This leaves 26 nuclear TDEs from eROSITA.

Another relevant comparison sample is the quiescent WBHs detected through their dynamical signatures. We searched the literature for individual reported detections. \citet{Taylor2025} presented a recent compilation of dynamically detected WBHs in UCDs in galaxy clusters: M60-UCD1, M59-UCD3, M59cO, VUCD3, Fornax UCD3, and UCD736 \citep{Seth2014,Ahn2017,Ahn2018,Afanasiev2018,Taylor2025}. Two further massive stellar clusters in M31, likely stripped nuclei, also host dynamically detected black holes: G1 and B023-G078 \citep{Gebhardt2005,Pechetti2022}. Finally, in the Milky Way's most massive stellar cluster, $\omega$~Centauri, itself likely a stripped nucleus \citep{Bekki2003}, a black hole with a lower mass limit of $\gtrsim 8000\,M_\odot$ has been reported by \citet{Haberle2024}.

Following our nomenclature, the UCD or stellar cluster associated with a dynamically detected WBH is called the stellar counterpart, and the galaxy in whose halo it lies is the parent galaxy. Several sources share the same parent, and most lie in the Virgo or Fornax clusters. We use the same approach to measure the parent $M_{\star}$ and halo masses and radii, the $M_{\star,\mathrm{scp}}$, and the radial distance to the parent. The dynamically detected sample is given in Table~\ref{tab:dyn}.

\begin{figure}[tbp]
    \includegraphics[width=\linewidth]{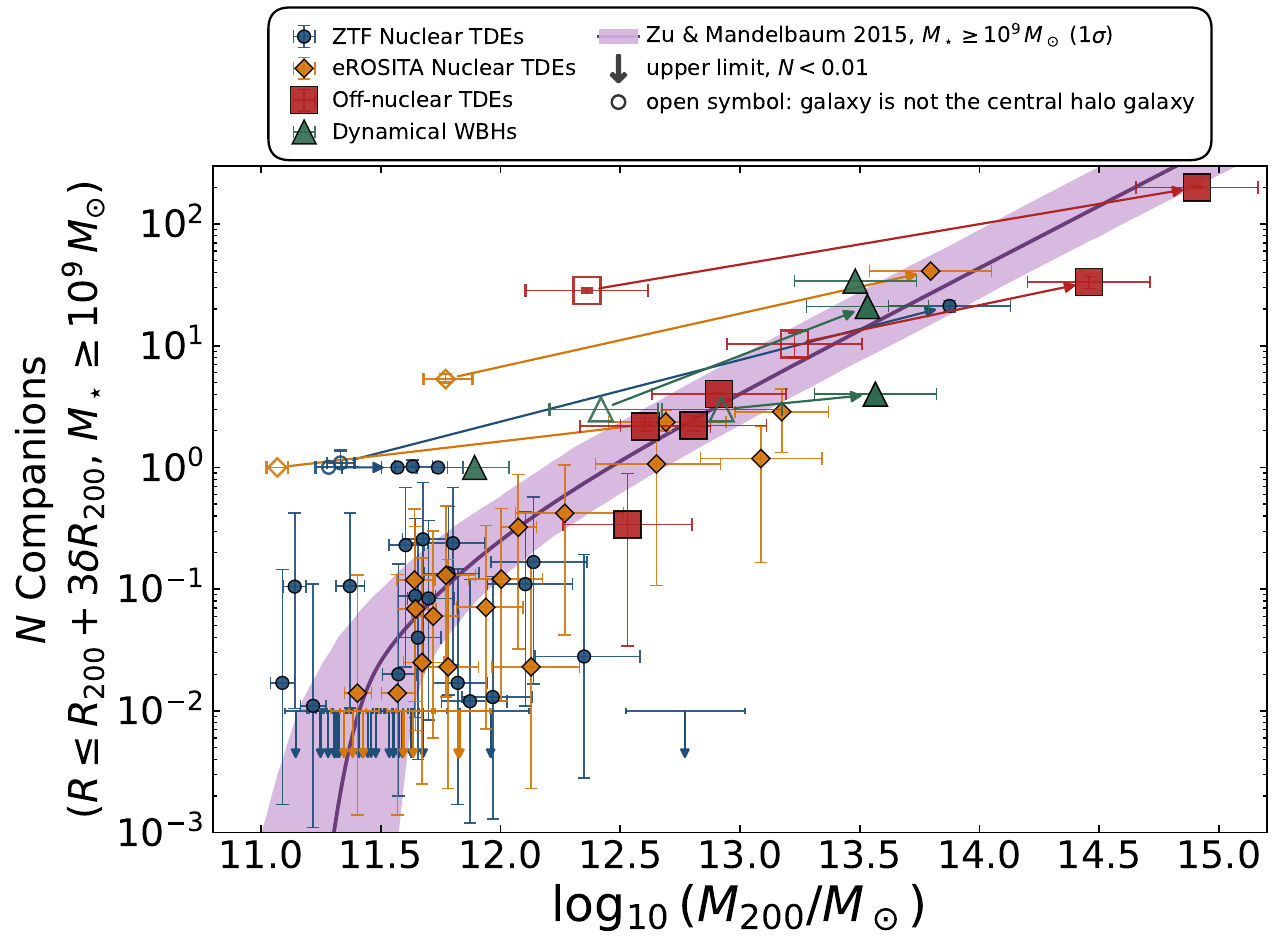}
   \caption{Number $N$ of spectroscopic and photometric companions with $M_\star \ge 10^{9}\,M_\odot$ against halo mass. Blue circles are ZTF nuclear TDEs, orange diamonds are eROSITA nuclear TDEs, red squares are off-nuclear TDEs, and green triangles are dynamically detected WBHs. The curve and band are the mean satellite occupation and its scatter from \citet{ZuMandelbaum2015}. Open symbols mark galaxies that are not the central galaxy of their halo; the arrow runs from that galaxy to the central galaxy and its $N$. Downward arrows are upper limits, $N<0.01$. $N$ can be fractional, because the probability that a photometric redshift satisfies the companion criteria can be $<1$ (see \S\ref{sec:mass_env}).}
    \label{fig:env}
\end{figure}

\subsection{\texttt{ROMULUS} Simulation}
\label{sec:romulus}

We summarize here the aspects of MBH modeling in the
\texttt{ROMULUS} simulations most relevant to this work; full technical details
are provided in \citet{Tremmel2017,Tremmel2019,Ricarte2021a}. The simulation suite
consists of \texttt{ROMULUS25}, a $(25\,\mathrm{Mpc})^{3}$ cosmological volume, and
\texttt{ROMULUSC}, a higher-resolution zoom-in simulation of a galaxy cluster with
$M_{200} \sim 10^{14}\,M_\odot$. Both were run using the Tree+SPH code
\texttt{ChaNGa} \citep{Menon2015,Wadsley2017}, with dark matter and gas particle
masses of $3.4\times10^{5}\,M_\odot$ and $2.1\times10^{5}\,M_\odot$, respectively.

The simulations add a correction for the dynamical friction from encounters
inside the gravitational softening length (a spline softening of 350~pc).
The correction integrates the formula of \citet{Chandrasekhar1943}
from the 90-degree deflection radius out to the softening length, supplying
only the drag missing from the resolved gravity \citep{Tremmel2015}. With
that force included, the simulations follow MBHs as they orbit freely within
their host halos, rather than artificially pinning them to galaxy centers. MBH
mergers are permitted once pairs become gravitationally bound within two
softening lengths, corresponding to $\sim0.7$~kpc.

This work directly adopts the
\texttt{ROMULUS} post-processing analysis presented in \citet{Ricarte2021a}. In
that study, halos were identified using the \textsc{Amiga} halo finder
\citep{Knollmann2009}, which was used to measure halo masses, radii, and centers.
MBHs were assigned to halos or substructures based on gravitational binding, and
halo centers were determined with an accuracy limited by the gravitational
softening length of 350 pc. In \citet{Ricarte2021a}, MBHs were classified as “central” if located within
0.7 kpc of the halo center and as “wandering” otherwise. This definition allows
for multiple central MBHs within a single halo and permits even the most massive
or luminous MBH to be classified as wandering. The adopted threshold reflects
uncertainties in MBH dynamics and center determination and corresponds to the
scale below which MBH mergers are allowed. Beyond this radius, MBH orbits are
well resolved and not artificially close to merging.

Here we explicitly use the numerical results describing the low-redshift ($z\sim0.05$) WBH population reported by \citet{Ricarte2021a}. These results are discussed in detail in the next section, but are summarized here for convenience: (i) the mean number of WBHs ($\langle N_{\rm WBH}\rangle$) per halo increases approximately linearly with halo virial mass $M_{200}$ \citep[Figure~2 of][]{Ricarte2021a}, reflecting the cumulative impact of hierarchical mergers and satellite accretion; (ii) WBHs span a broad range of halo-centric distances ($R/R_{200}$), with a roughly Gaussian distribution extending from the halo nucleus to near the virial radius. Specifically, we use the radial distribution of WBHs in the halo-mass bins relevant to our study, $\log_{10}(M_{200}/M_{\odot}) \sim 12.5$--$13$ and $13$--$13.5$, shown in Figure~6 of \citet{Ricarte2021a}.

Another result, which we do not use directly but which is relevant, is that at $z=0.05$ most \texttt{ROMULUS} WBHs show no stellar counterpart detectable by the post-processing halo finder \citep[their Figure~9]{Ricarte2021a}. This does not mean that the WBHs are truly free-floating, but only that their original galaxies have lost most of their resolved stellar disk and halo. This general result also seems to hold in the \texttt{ASTRID} cosmological simulation, where minor mergers tidally strip the satellite and leave its black hole wandering \citep{Weller2026}. A dense nuclear stellar cluster would not be resolved by \texttt{ROMULUS}, and higher-resolution simulations show that such a cluster can survive tidal stripping during infall \citep{Bekki2001,Pfeffer2013}.

\section{Comparing Populations}\label{sec:results_obs}
In this section, we present an observational comparison of some demographic properties of off-nuclear TDEs and the samples discussed in \S\ref{sec:comp_samp}.

\subsection{Galaxy Masses and Environment}\label{sec:mass_env}

From Table~\ref{tab:hosts} it is immediately apparent, as first noted by \citet{Yao2025}, that the parent galaxies of off-nuclear TDEs share remarkably similar global properties. All six occur in massive, early-type galaxies, with $M_{\star}$ confined to $10.7 \lesssim \log_{10}(M_\star/M_\odot) \lesssim 11.1$. For a sample this small, the agreement is striking, and it suggests a physical origin rather than a coincidence. We shall examine that origin in \S\ref{sec:results_sim}; here we compare these hosts with nuclear TDEs and with dynamically detected WBHs.

Galaxies of this mass and type are not where most nuclear TDEs are found \citep{Hammerstein_21,Yao2023}. Early-type galaxies here typically have $\sigma_\star \gtrsim 200~\mathrm{km~s^{-1}}$, and the $M_\bullet$--$\sigma_\star$ relation then gives $M_\bullet \gtrsim 10^{8}\,M_\odot$ \citep{Kormendy2013}. The intrinsic nuclear TDE rate is strongly suppressed at these $M_{\bullet}$: the per-galaxy rate declines with $M_\bullet$ \citep{Stone2016,Hannah2025}; above the Hills mass, stars are swallowed whole for a wide range of stellar mass and spin \citep{hills,Kesden12_maximum_Mbh,Mummery_vanVelzen2024}\footnote{The tidal radius grows only as $r_{\rm T} \propto M_{\bullet}^{1/3}$, while the event horizon grows as $r_{\rm H} \propto M_{\bullet}$.}; and galaxies this massive are rare. The three effects together produce a super-exponential drop in the intrinsic rate \citep[][Karmen et al., in prep.]{vanVelzen2018,Mummery_vanVelzen2024}. Flux-limited optical surveys still recover a few of them, because the early-time peak luminosity scales approximately linearly with $M_{\bullet}$, $L_{\rm peak,opt} \propto M_{\bullet}$ \citep{Mummery2024,Mummery2025Calorimetry}, which compensates for the low rate. Those objects are the high-mass tail in Fig.~\ref{fig:mass-color}.

Fig.~\ref{fig:mass-color} shows the comparison directly. The gray background is the volume-corrected distribution of the Sloan Digital Sky Survey (SDSS) main galaxy sample \citep{Strauss2002}. The nuclear TDEs from ZTF and eROSITA (\S\ref{sec:comp_samp}) span $9 < \log_{10}(M_\star/M_\odot) < 11$, with a median $\log_{10}(M_\star/M_\odot) \sim 10$, and are overrepresented in the green valley \citep{Law-Smith_17,Hammerstein_21,Yao2023}. The off-nuclear parents lie together at the red, high-mass end, and the cumulative distributions separate the two samples in both mass and color.

The dynamically detected WBHs on the same figure (\S\ref{sec:comp_samp}; Table~\ref{tab:dyn}) follow the off-nuclear parents: red galaxies, the same narrow range in mass, and closely similar cumulative distributions. The two samples are consistent with one parent population, seen in accretion as off-nuclear TDEs and in quiescence through dynamical detection, with the selection function set by the discovery method. The Milky Way and $\omega$~Cen are the exceptions. The Milky Way has a much lower $M_{\star}$ and is a green-valley galaxy, a product of a discovery method that is extremely different from that of the rest of the dynamical sample.

A characteristic of high-mass early-type galaxies is that they inhabit groups and clusters rather than the field \citep{Dressler1980,PostmanGeller1984,Kauffmann2004}, as expected if they assemble hierarchically inside massive dark-matter halos \citep{WhiteRees1978,DeLucia2006}. Off-nuclear TDEs and dynamically detected WBHs, which occur in galaxies of this kind, should therefore favor such environments, while the lower-mass hosts of nuclear TDEs should more often be isolated. To test this, we measure for each parent or host galaxy the number $N$ of companions with $M_\star \ge 10^{9}\,M_\odot$.

The measurement is described in detail in Appendix~\ref{app:3}. A companion $i$ enters $N$ when its $M_{\star}$, recession velocity, and projected separation satisfy
\begin{itemize}
\item $M_{\star,i} \ge 10^{9}\,M_\odot$,
\item $|cz - cz_i| \leq 5\sigma_{\rm los}$,
\item $R_p \leq R_{200} + 3\delta R_{200}$.
\end{itemize}

Here $cz$ and $cz_i$ are the recession velocities of the host and of galaxy $i$, and $R_p$ is their projected separation. $\delta R_{200}$ is the upper $1\sigma$ uncertainty on $R_{200}$, so the search aperture is $R_{200}$ extended by three times that error. The velocity scale is the halo line-of-sight dispersion, $\sigma_{\rm los} = V_{200}/\sqrt{2}$, with $V_{200} = (GM_{200}/R_{200})^{1/2}$. These conditions indicate that the companion is gravitationally bound to the halo. In the computation of $N$, a spectroscopically confirmed companion satisfying the three conditions adds 1 to $N$. A photometric redshift adds the probability that the galaxy lies inside the velocity window and above $10^{9}\,M_\odot$. The resulting $N$ is therefore not restricted to integer values, and it quantifies distinct environments.

In Fig.~\ref{fig:env} we show $N$ for the comparison samples as a function of halo mass $M_{200}$, together with the mean number of $M_\star \geq 10^{9}\,M_\odot$ satellites expected in a halo of that mass from the occupation model of \citet{ZuMandelbaum2015}. Off-nuclear TDEs and dynamically detected WBHs lie in much denser environments than nuclear TDEs. Five of the six off-nuclear hosts have $N \geq 1$, the exception being AT2023mfm, and all five parents in the dynamical WBH sample have $N \geq 1$. Of the nuclear TDEs, only 4 of 45 ZTF hosts and 5 of 26 eROSITA hosts do.

For some of the sources, computing $N$ showed that the parent was not the central galaxy of the halo, that is, not the most massive galaxy in it. We then place the halo center on the most massive galaxy, adopt its $R_{200}$, and recompute $N$. In Fig.~\ref{fig:env} these systems are shown as open symbols, with an arrow pointing to the central galaxy and its $N$. Two off-nuclear parents are of this kind: ESO 243-49 (HLX-1) and AT2025abcr. Both are massive early-type galaxies, but they are not the centrals of their halos: unlike the other off-nuclear hosts with $N > 1$, which lie in groups, these two sit in extremely massive galaxy clusters, with $M_{200} \gtrsim 10^{14.5}\,M_\odot$.

\subsection{Radial Distribution and Stellar Counterparts}

\begin{figure}[tbp]
    \centering
    \includegraphics[width=\linewidth]{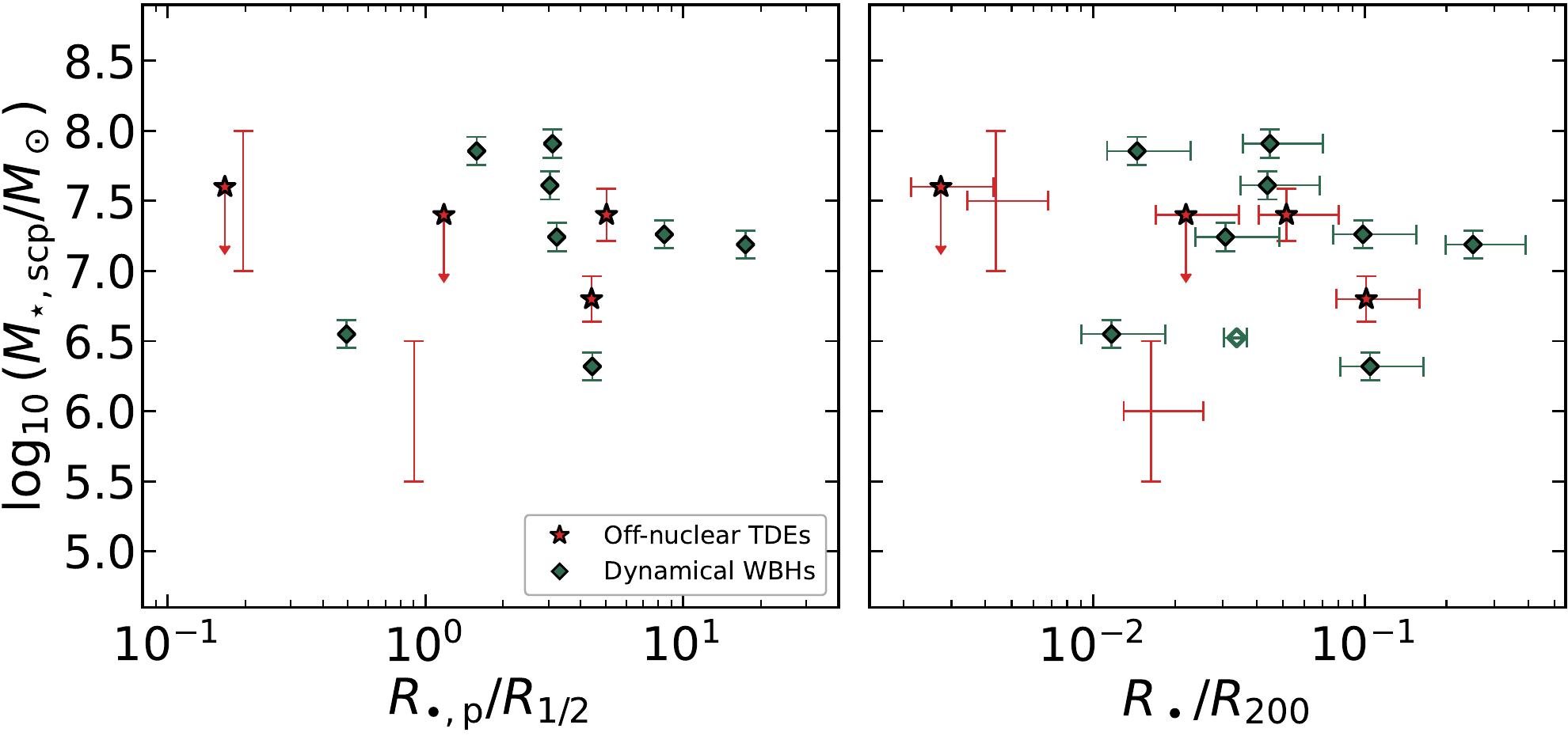}
    \caption{$M_{\star,\mathrm{scp}}$ against projected offset in units of the parent half-light radius (left) and deprojected offset in units of $R_{200}$ (right). Red stars are off-nuclear TDEs and green diamonds are dynamically detected WBHs. $\omega$~Cen is the open diamond; it has no sky-projected offset and appears only on the right. Vertical arrows are upper limits, and vertical bars without a symbol are the possible mass ranges for post-TDE detected counterparts.}
    \label{fig:R_scp}
\end{figure}

If the dynamically detected WBHs are the quiescent counterparts of the transient, accreting population traced by off-nuclear TDEs---a reading already supported by the similar parent galaxies and environments of the previous section---their offsets from the host nucleus and their $M_{\star,\mathrm{scp}}$ should be comparable, although those differences originate from the distinct selection and discovery methods. Figure~\ref{fig:R_scp} shows both quantities: the projected offset $R_{\bullet,\mathrm{p}}/R_{1/2}$ and the deprojected offset $R_{\bullet}/R_{200}$, against $M_{\star,\mathrm{scp}}$.

Dynamical WBHs are found, on average, at larger projected separations. Excluding $\omega$~Cen, for which a sky-projected offset from the Galactic Center is not well defined,\footnote{The galactocentric radius of $\omega$~Cen is a measured three-dimensional distance, so $R_{\bullet,\mathrm{p}}$ is not defined in the same way as for the external galaxies.} seven of the eight sources lie at $R_{\bullet,\mathrm{p}}/R_{1/2}>1$. The same threshold splits the off-nuclear TDEs in half. Off-nuclear TDEs therefore reach WBHs inside one stellar half-light radius, a region that dynamical studies have so far barely recovered.

The stellar counterparts of the dynamical sources are also more massive. Six of the nine dynamical WBHs, now including $\omega$~Cen, have $M_{\star,\mathrm{scp}}\gtrsim10^{7}\,M_{\odot}$. Of the six off-nuclear TDEs, two have $M_{\star,\mathrm{scp}}\gtrsim10^{7}\,M_{\odot}$, two lie below $10^{7}\,M_{\odot}$, and two have no detected counterpart, with limits of a few $\times10^{7}\,M_{\odot}$.

While we are not exploring $M_{\bullet}$ in this paper, it is reasonable to expect that dynamical detections should favor higher $M_{\bullet}$. A dynamical measurement requires the radius of influence to be spatially resolved, and, at fixed stellar velocity dispersion, that radius scales linearly with $M_{\bullet}$, $R_{\mathrm{infl}}=GM_{\bullet}/\sigma_{\star}^{2}$.

\section{Observables vs. \texttt{ROMULUS}}\label{sec:results_sim}
\subsection{Parent Galaxy Preference}
\label{sec:parent_pref}

As shown in \S\ref{sec:results_obs}, the off-nuclear TDEs all lie in early-type, quiescent galaxies, and all lie in a narrow range of relatively high $M_{\star}$, $10.7 \lesssim \log_{10}(M_\star/M_\odot) \lesssim 11.1$. These masses are higher than those of typical nuclear TDE hosts, but not as high as those of BCGs, so the galaxies sit between the two populations. With only a handful of sources, a distribution this uniform in mass and color is unlikely to be accidental, and it should trace a physical origin.

In this section we show that the preference for this class of galaxies arises naturally from the demographics of WBHs predicted by \texttt{ROMULUS}, once combined with the observed local galaxy population. To this end, we compute the volumetric density of WBHs as a function of $M_{\star}$, $\phi_{\rm WBH}(M_\star)$.

\begin{figure*}
    \centering
    \includegraphics[width=\linewidth]{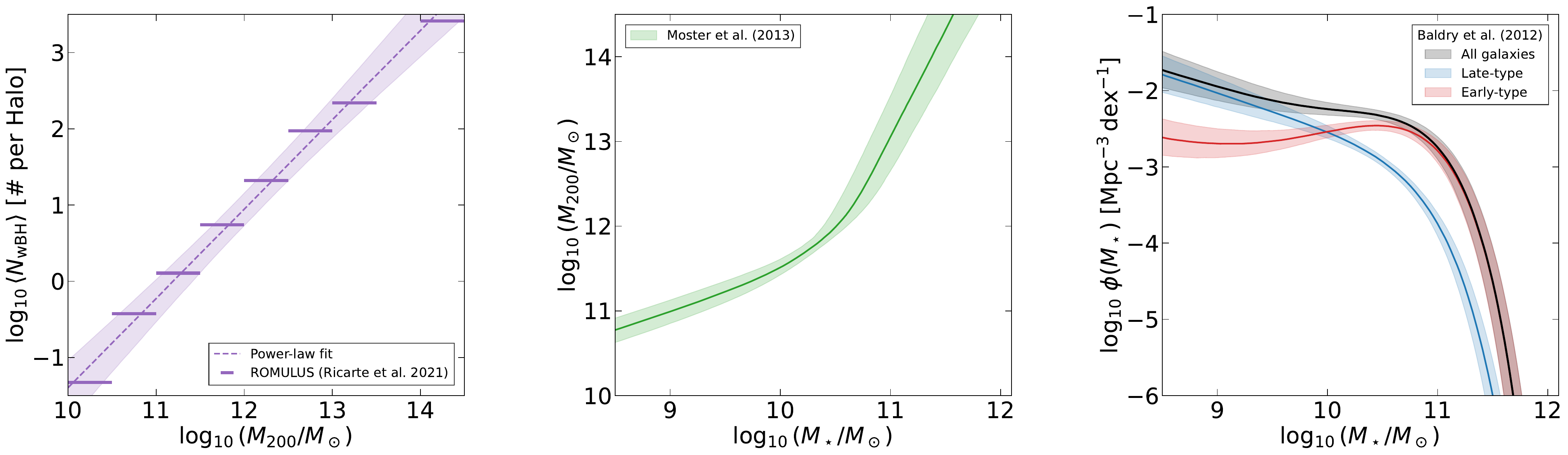}
    \caption{\textit{Left:} Mean number of WBHs per halo as a function of halo virial mass at $z\simeq 0.05$ in the \texttt{ROMULUS} simulations \citep{Ricarte2021a}. Purple points show the binned simulation results, the dashed line shows the best-fitting power-law relation, and the shaded region indicates the $1\sigma$ uncertainty.
    \textit{Middle:} Stellar-to-halo mass relation at $z\simeq 0$ from \citet{Moster2013}, mapping $M_{200}$ to $M_{\star}$.
    \textit{Right:} Local GSMF from \citet{Baldry2012}, for all galaxies (black) and divided into late-type (star-forming, blue) and early-type (quenched, red) galaxies.
    These three ingredients combine to give the volumetric density of WBHs in Fig.~\ref{fig:WBH_density}.}
    \label{fig:ingredients}
\end{figure*}

As discussed in \S\ref{sec:romulus} and shown in the left panel of Fig.~\ref{fig:ingredients}, the mean number of WBHs per halo in \texttt{ROMULUS} scales approximately linearly with halo virial mass,
\begin{equation}
\log_{10}\langle N_{\rm WBH} \rangle
= A\,
\log_{10}\!\left(\frac{M_{200}}{10^{12}\,M_\odot}\right)
+ B,
\label{eq:Nwbh}
\end{equation}
where $A = 1.17\pm 0.17$ and $B = 0.95\pm 0.22$.\footnote{Only the average number of WBHs per bin of $0.5$~dex in halo mass is given by \citet{Ricarte2021a}, so the uncertainties reflect the scatter associated with that bin size.}

The quasi-linear $\langle N_{\rm WBH}\rangle$--$M_{200}$ relation is a key ingredient of this result. The absolute normalization depends on the black-hole seeding prescription adopted in \texttt{ROMULUS}, but the slope is expected to be a robust feature of hierarchical structure formation. Other tracers of that assembly show the same scaling. For example, the number of resolved satellite galaxies \citep{Zehavi2011,ZuMandelbaum2015} and the number of globular clusters \citep{Harris2013,Hudson2014,Harris2015,BurkertForbes2020} both increase roughly linearly with dark-matter halo mass. As with those tracers, the accumulation of WBHs is driven primarily by the merger and accretion history of the central halo, rather than by the details of early black-hole formation. Alternative seeding models may shift the normalization of $\langle N_{\rm WBH}\rangle$, or add scatter at low halo mass, where the seeding prescriptions have their largest effect \citep[e.g.,][]{Untzaga2024,Weller2026b}. They are not expected to change the strong, monotonic increase of WBH abundance with halo mass, which is set by hierarchical merging and satellite accretion over cosmic time\footnote{Even if the seeding prescription slightly affects the correlation, Appendix~\ref{app:c} shows that any $\langle N_{\rm WBH}\rangle \propto M_{200}^{A}$ with $0.9 \leq A \leq 1.3$, still monotonic and quasi-linear, would explain the observed parent-galaxy mass preference.}.

To express this relation in terms of $M_{\star}$, we adopt the stellar-to-halo mass relation (SHMR) of \citet{Moster2013},
\begin{equation}\label{eq:SHMR}
M_\star(M_{200}) =
2\,N\,M_{200}
\left[
\left( \frac{M_{200}}{M_1} \right)^{-\beta}
+
\left( \frac{M_{200}}{M_1} \right)^{\gamma}
\right]^{-1},
\end{equation}
with best-fit parameters at $z \simeq 0$ of $\log_{10}(M_1/M_\odot)=11.6\pm0.2$, $N=0.035\pm0.010$, $\beta=1.38\pm0.15$, and $\gamma=0.61\pm0.06$. This mapping is shown in the middle panel of Fig.~\ref{fig:ingredients}.

The abundance of galaxies as a function of $M_{\star}$ is taken from the local GSMF of \citet{Baldry2012}. Written as a double Schechter function, it is
\begin{equation}
\begin{aligned}
\phi_{\rm gal}(M_\star)
&=
\ln(10)\,e^{-M_\star/M^\ast} \\
&\quad\times
\left[
\phi_1^\ast \left(\frac{M_\star}{M^\ast}\right)^{\alpha_1+1}
+
\phi_2^\ast \left(\frac{M_\star}{M^\ast}\right)^{\alpha_2+1}
\right],
\end{aligned}
\label{eq:gsmf}
\end{equation}
with $\phi_{\rm gal}$ in $\mathrm{Mpc}^{-3}\,\mathrm{dex}^{-1}$ and parameters $\log_{10}(M^\ast/M_\odot)=10.66\pm0.02$, $\phi_1^\ast=(3.96\pm0.34)\times10^{-3}\,\mathrm{Mpc}^{-3}$, $\alpha_1=-0.35\pm0.10$, $\phi_2^\ast=(0.79\pm0.07)\times10^{-3}\,\mathrm{Mpc}^{-3}$, and $\alpha_2=-1.47\pm0.05$. The GSMF is shown in the right panel of Fig.~\ref{fig:ingredients}.

The volumetric density of WBHs per dex of $M_{\star}$ is then
\begin{equation}
\phi_{\rm WBH}(M_\star)
=
\phi_{\rm gal}(M_\star)\,
\langle N_{\rm WBH}[M_{200}(M_\star)] \rangle .
\label{eq:phi_wbh}
\end{equation}
Equation~\ref{eq:phi_wbh} captures the two effects that shape the demographics. Low-mass galaxies, $\log_{10}(M_\star/M_\odot)\lesssim 10$, dominate the galaxy number density but host on average $\langle N_{\rm WBH}\rangle \lesssim 1$ WBH. Very massive galaxies, $\log_{10}(M_\star/M_\odot)\gtrsim 11.3$, host hundreds to thousands of WBHs each, but are exponentially rare. Their product therefore peaks at intermediate-to-high $M_{\star}$, where galaxies are both sufficiently common and individually rich in WBHs.

\begin{figure*}[t]
\centering
\includegraphics[width=0.7\textwidth]{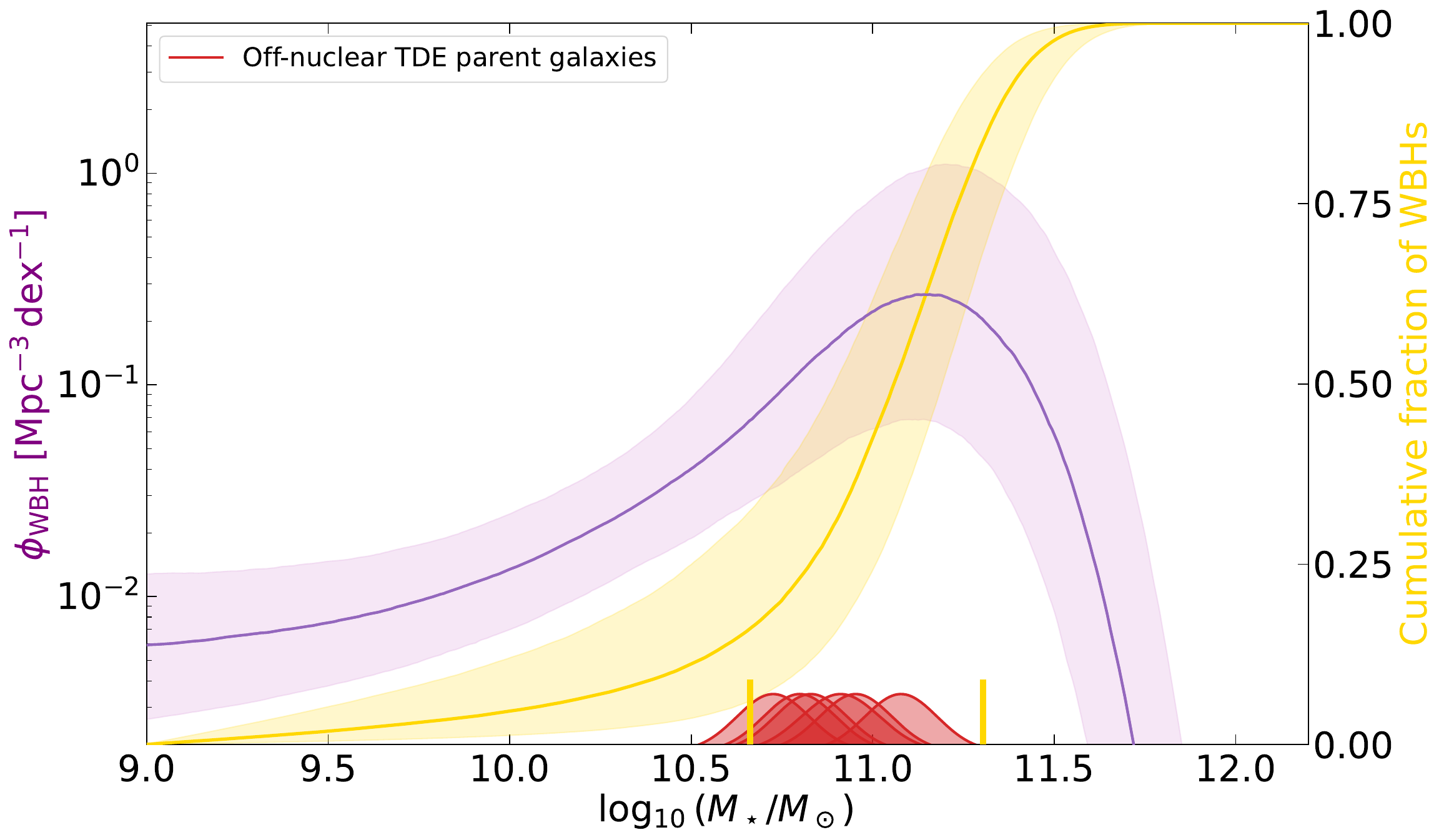}
\caption{
Volumetric density of WBHs as a function of $M_{\star}$ in the local Universe.
The purple curve and shaded band show $\phi_{\rm WBH}(M_\star)$, in $\mathrm{Mpc}^{-3}\,\mathrm{dex}^{-1}$.
The yellow curve is the cumulative fraction of WBHs, and the yellow ticks mark the 16th and 84th percentiles of that distribution.
The density peaks at $M_\star\sim 10^{11.1}\,M_\odot$, with $10.65 \lesssim \log_{10}(M_\star/M_\odot) \lesssim 11.30$ enclosing 68\% of the population.
The red kernels are the off-nuclear TDE parent galaxies.}
\label{fig:WBH_density}
\end{figure*}

The resulting $\phi_{\rm WBH}(M_\star)$ is the purple curve in Fig.~\ref{fig:WBH_density}. It peaks sharply (note the log scale) at $\log_{10}(M_\star/M_\odot)\sim 11.1$. The yellow cumulative distribution is narrow: most (central 68\%) WBHs in the local Universe reside in galaxies with $10.65 \lesssim \log_{10}(M_\star/M_\odot) \lesssim 11.30$.

Of course, the intrinsic off-nuclear TDE rate is not set by $\phi_{\rm WBH}$ alone. In general form we can write it as

\begin{equation}\label{eq:rate_int}
\rho^{\rm off-nuclear}_{\rm int}(M_\star, M_{\bullet})
= \phi_{\rm WBH}(M_\star)\,r(M_{\bullet})\,\psi(M_\star, M_{\bullet}),
\end{equation}

The rate per black hole, $r(M_{\bullet})$, is in units of $\mathrm{yr}^{-1}$ and does not depend on $M_\star$. The function $\psi(M_\star, M_{\bullet})$ is the distribution of $M_{\bullet}$ in a parent of mass $M_\star$, normalized so that $\int \psi(M_\star, M_{\bullet})\,dM_{\bullet} = 1$.

A survey that finds TDEs at wavelength $\lambda$, to a finite depth, will, however, be affected by the survey selection and by the TDE emission at that wavelength as a function of $M_{\bullet}$, and it will therefore see an observational distribution, which takes the general form

\begin{multline}\label{eq:rate_obs}
\rho^{\rm off-nuclear}_{\rm obs,\lambda}(M_\star, M_{\bullet})
= \\ \phi_{\rm WBH}(M_\star)\,r(M_{\bullet})\,\psi(M_\star, M_{\bullet}) \\
\times
\int k_{\lambda}\!\left(L_{{\rm peak},\lambda}\mid M_{\bullet}\right)\,
\mathcal{E}_{\lambda}\!\left(L_{{\rm peak},\lambda}\right)\,dL_{{\rm peak},\lambda},
\end{multline}

where $L_{{\rm peak},\lambda}$ is the peak luminosity, $k_{\lambda}(L_{{\rm peak},\lambda}\mid M_{\bullet})$ is its probability distribution at fixed $M_{\bullet}$, normalized to unity, and $\mathcal{E}_{\lambda}(L_{{\rm peak},\lambda})$ is the survey exposure: the comoving volume and time over which a flare of that luminosity is detected and classified in the survey. 

The same expression can be constructed for a general survey at a given wavelength, for example an optical survey such as ZTF or an X-ray survey such as eROSITA. In the optical, $k_{\lambda}$ can be built from the empirical linear relation $L_{{\rm peak,opt}}\propto M_{\bullet}$ \citep[][Karmen et al., in prep.]{Mummery2024,Mummery2025Calorimetry}. In X-rays, $L_{{\rm peak,X-ray}}(M_{\bullet})$ is a more complex and nonlinear function of $M_{\bullet}$, but it can be constructed from time-dependent disk-theory simulations \citep{Mummery_vanVelzen2024}.

The present off-nuclear sample is very heterogeneously selected. Events were found serendipitously with non-survey X-ray instruments, in one X-ray survey, and in one optical survey. Each channel has its own peak-luminosity distribution and its own survey exposure, so $k_{\lambda}\!\left(L_{{\rm peak},\lambda}\mid M_{\bullet}\right)$ and $\mathcal{E}_{\lambda}\!\left(L_{{\rm peak},\lambda}\right)$ cannot be written for the sample as a whole. The wanderer mass distribution $\psi(M_\star, M_{\bullet})$ can be measured only in a cosmological simulation that, unlike \texttt{ROMULUS}, does not impose a uniform and high seed mass. \texttt{ASTRID} is an example, but to our knowledge this has not been done. We therefore cannot evaluate the full form of Eq.~\ref{eq:rate_int} or Eq.~\ref{eq:rate_obs}.

The observed parent galaxies nonetheless seem to follow $\phi_{\rm WBH}(M_\star)$. The observed sample is a sum of many realizations of Eq.~\ref{eq:rate_obs}, one for each discovery survey or instrument, yet the parents lie at $10.7 \lesssim \log_{10}(M_\star/M_\odot) \lesssim 11.1$, inside the central 68\% of $\phi_{\rm WBH}(M_\star)$, $10.65 \lesssim \log_{10}(M_\star/M_\odot) \lesssim 11.30$. For the wanderer masses relevant here, a few $\times 10^{4} \lesssim M_{\bullet}/M_\odot \lesssim$ a few $\times 10^{6}$, it appears that the supply of WBHs therefore dominates the dependence of $\rho^{\rm off-nuclear}_{\rm obs,\lambda}(M_\star)$ on parent mass. The remaining terms in Eq.~\ref{eq:rate_obs} add scatter at fixed $M_\star$, but they do not move the observed events into halos of a different mass. If they did, the peak of $\phi_{\rm WBH}$ would not coincide with the observed parent masses.

In this spirit, it is therefore reasonable, to first order, to assume
\begin{equation}\label{eq:approximation}
\rho^{\rm off-nuclear}_{\rm obs,\lambda}(M_\star)
\propto
\rho^{\rm off-nuclear}_{\rm int}(M_\star)
\propto
\phi_{\rm WBH}(M_\star),
\end{equation}
which simply implies that, for any survey, the observed rate as a function of parent-galaxy mass is dominated by the volumetric supply of WBHs at that mass, while observational effects and $M_{\bullet}$-dependent effects only add scatter and remain subdominant.

This approximation also allows us to compare the nuclear and off-nuclear intrinsic rates as a function of host or parent $M_{\star}$. Given their distinct shapes, this comparison should present an interesting effect: because the off-nuclear rate peaks at high $M_{\star}$ and the nuclear rate falls steeply toward high host mass, above a characteristic $M_{\star}$ the off-nuclear rate should exceed the nuclear rate.

\begin{figure}[tbp]
\centering
\includegraphics[width=\linewidth]{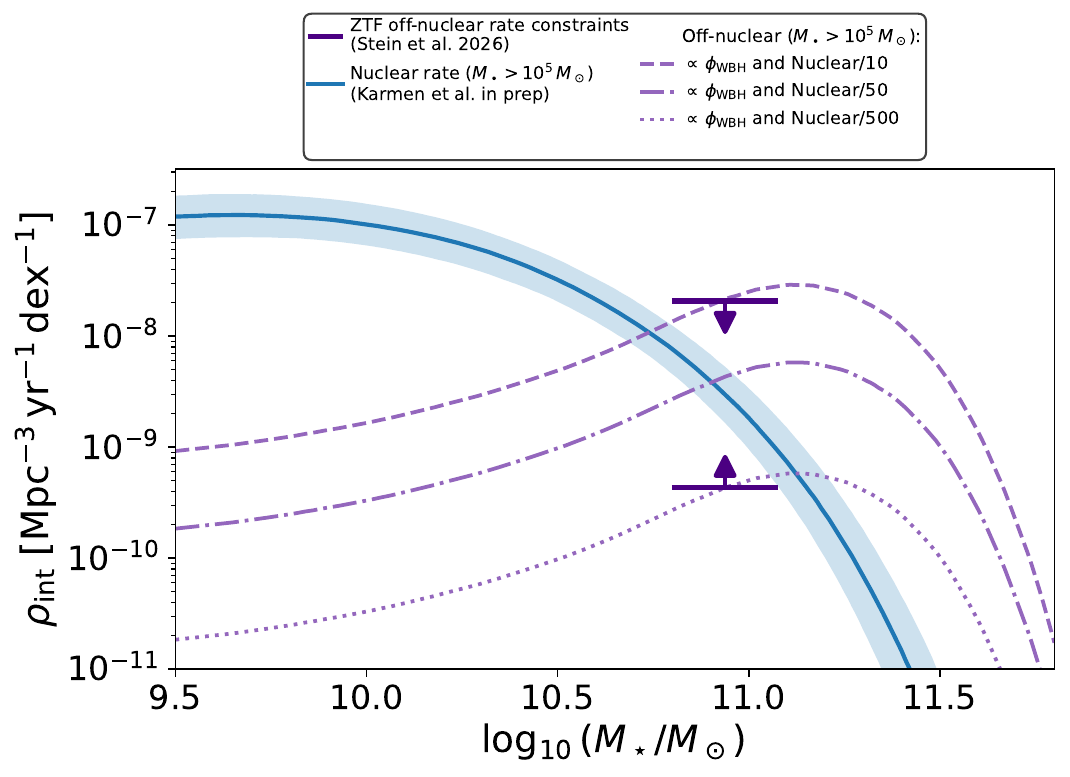}
\caption{
Intrinsic TDE rate per dex of $M_{\star}$ for black holes with $M_{\bullet}>10^{5}\,M_\odot$.
The blue curve is the nuclear rate as a function of host $M_{\star}$, from Karmen et al.\ (in prep.).
The purple curves are possible off-nuclear rates, under the assumption that the rate is, to leading order, proportional to $\phi_{\rm WBH}(M_{\star})$ and that its integrated rate is a fraction $f$ of the nuclear rate. The curves adopt $f=1/10$, $1/50$, and $1/500$.
The purple limits are the upper and lower limits on the off-nuclear TDE rate from \citet{Stein2026}, drawn across the $M_{\star}$ range of the ZTF off-nuclear parent galaxies.}
\label{fig:nuc_vs_off-nuc}
\end{figure}

In Fig.~\ref{fig:nuc_vs_off-nuc} we show the intrinsic nuclear TDE rate for black holes with $M_{\bullet}>10^{5}\,M_\odot$,\footnote{This is the lower $M_{\bullet}$ limit that ZTF can constrain \citep[][Karmen et al., in prep.]{Yao2023,Mummery_vanVelzen2024}.} expressed as a function of host $M_{\star}$, $\rho^{\rm nuclear}_{\rm int}(M_\star, M_{\bullet}>10^{5}\,M_\odot)$, as derived by Karmen et al.\ (in prep.). With the assumption of Eq.~\ref{eq:approximation}, the off-nuclear rate can be expressed as

\begin{multline}
\rho^{\rm off-nuclear}_{\rm int}(M_\star)\Big|_{M_{\bullet}>10^{5}\,M_{\odot}}=\\ \mathcal{F}\times
\frac{\phi_{\rm WBH}(M_\star)}
{\displaystyle\int \phi_{\rm WBH}(M_\star')\,{\rm d}\log_{10} M_\star'},
\end{multline}

where $\mathcal{F}$ is defined so that the integrated off-nuclear rate is a fraction $f$ of the nuclear rate \citep[which at $M_{\bullet}>10^{5}\,M_{\odot}$ is well known,][Karmen et al.\, in prep]{Yao2023},
\begin{equation}
\mathcal{F}
=
f\int_{M_{\bullet}>10^{5}\,M_{\odot}}
\rho^{\rm nuclear}_{\rm int}(M_{\bullet})\,{\rm d}\log_{10} M_{\bullet}.
\end{equation}

The curves in Fig.~\ref{fig:nuc_vs_off-nuc} adopt $f=1/10$, $1/50$, and $1/500$. We choose these values because the largest and the smallest are the upper and lower limits on the off-nuclear-to-nuclear rate derived by \citet{Stein2026} from ZTF selection, $9.6\%$ and $0.2\%$ of the nuclear rate, respectively. From these assumptions and this constraint, the off-nuclear TDE rate should surpass the nuclear rate at high $M_{\star}$.

To conclude, the observed preference of off-nuclear TDEs for massive galaxies, but not the most massive BCGs, follows simply from the demographics of WBHs. Even if the probability that an individual WBH produces a TDE depends on $M_{\bullet}$ \citep{Hannah2025,Stone2016}, and a given survey preferentially selects a distinct range of $M_{\bullet}$ \citep[e.g., ZTF cannot detect $M_{\bullet} < 10^{5}\,M_{\odot}$ given its depth and the $L_{\rm peak,opt} \propto M_{\bullet}$ relation,][]{Mummery_vanVelzen2024,Ramsden2025}, the parent-galaxy distribution will still cluster around $M_\star\sim 10^{11}\,M_\odot$, simply because that is where the cosmic density of WBHs is largest and dominates the other effects. 

The same halo-mass scaling explains the distinct environments of nuclear and off-nuclear TDEs. Because $\langle N_{\rm WBH}\rangle$ increases with $M_{200}$, the parents that dominate $\phi_{\rm WBH}$ inhabit relatively massive dark-matter halos, and those halos contain more companion galaxies. Off-nuclear TDEs are therefore found in groups and clusters more often than nuclear TDEs, whose hosts lie at lower $M_{\star}$ and in lower-mass halos. The parent-galaxy preference is a demographic consequence of hierarchical structure formation. 

Furthermore, the early-type morphology of the parent galaxies follows the same reasoning. Over $10.7 \lesssim \log_{10}(M_\star/M_\odot) \lesssim 11.2$, early-type galaxies are $\sim$7 times more common than late-type galaxies, while star-forming late-type galaxies are already exponentially suppressed there (right panel of Fig.~\ref{fig:ingredients}). The morphological preference is therefore explained by the galaxy type distribution at the $M_{\star}$ where $\phi_{\rm WBH}(M_\star)$ is largest.

A clear prediction is that, as the sample grows, parent galaxies will continue to be drawn from $10.65 \lesssim \log_{10}(M_\star/M_\odot) \lesssim 11.30$. Galaxies with $\log_{10}(M_\star/M_\odot)\lesssim 10.5$ are very unlikely to host an off-nuclear TDE. Galaxies with $\log_{10}(M_\star/M_\odot)\gtrsim 11.3$ are rare, but their halos contain so many WBHs (potentially more than in \texttt{ROMULUS}, which includes only one massive cluster) that an off-nuclear TDE can still appear as the sample grows. WINGS~J1348$+$26 (\S\ref{sec:sample}), in Abell~1795, may already be such a case. Its BCG has $\log_{10}(M_\star/M_\odot)\approx 11.6$, although whether the TDE is off-nuclear with respect to that galaxy is ambiguous, as discussed in \S\ref{sec:sample}.

\subsection{Radial Distribution}
\label{sec:radial}

Another demographic property of the current off-nuclear TDE sample is the wide range of halo-centric radii at which these events occur. From Table~\ref{tab:hosts}, the six events span nearly two orders of magnitude in normalized radius, from $R_{\bullet}/R_{200}\simeq0.003$ to $R_{\bullet}/R_{200}\simeq0.1$, or equivalently from $R_{\bullet,\mathrm{p}}/R_{1/2}\simeq0.2$ to $R_{\bullet,\mathrm{p}}/R_{1/2}\simeq6$.

\begin{figure}
    \centering
    \includegraphics[width=\linewidth]{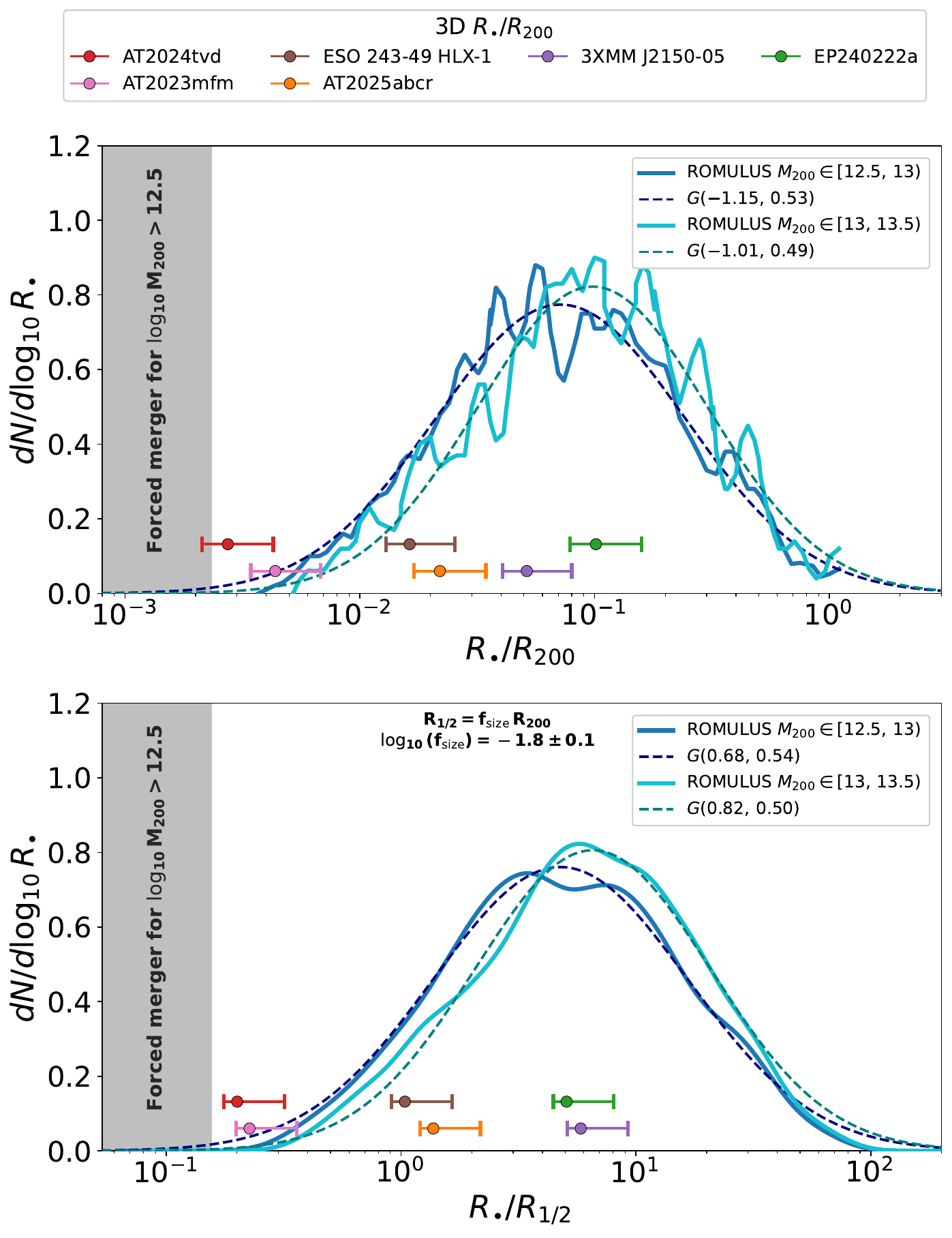}
\caption{
Radial distribution of WBHs in \texttt{ROMULUS} compared with the
deprojected locations of the off-nuclear TDEs. Solid curves
show $\mathrm{d}N/\mathrm{d}\log_{10}R$ for halos with
$\log_{10}(M_{200}/M_\odot)=12.5$--13 (blue) and 13--13.5 (cyan)
\citep{Ricarte2021a}. Dashed curves are Gaussian fits in $\log R_{\bullet}$. The gray band marks radii inside the forced-merger distance for halos above
$\log_{10}(M_{200}/M_\odot)=12.5$. The lower panel shows the same curves in units of $R/R_{1/2}$, adopting
$R_{1/2}=f_{\mathrm{size}}\,R_{200}$ with a lognormal $f_{\mathrm{size}}$ of
median $0.015$ and $\sigma_{\log f}=0.1$~dex. The data points there are
$R_{\rm \bullet}/R_{1/2}$ from Table~\ref{tab:hosts}. Points show the median
radius, with bars spanning the 16th--84th percentiles.
}
\label{fig:R_dist}
\end{figure}

The top panel of Fig.~\ref{fig:R_dist} compares these locations with the radial distribution of WBHs predicted by \texttt{ROMULUS} \citep[Figure~6 of][]{Ricarte2021a}. The solid curves show $\mathrm{d}N/\mathrm{d}\log_{10}R$ at $z=0.05$ for halos with $\log_{10}(M_{200}/M_\odot)=12.5$--13 (blue) and 13--13.5 (cyan), the halo masses most relevant to this sample. The dashed curves are the best-fit Gaussians in $\log_{10}(R_{\bullet}/R_{200})$, which describe the distribution well to leading order. The gray band marks the radii inside which black-hole pairs are forced to merge in \texttt{ROMULUS}, for every halo above $\log_{10}(M_{200}/M_\odot)=12.5$. The points are the median deprojected radii of the events in Table~\ref{tab:hosts}, with bars spanning the 16th--84th percentiles.

The lower panel of Fig.~\ref{fig:R_dist} shows the same distribution in units of the host half-light radius, $R_{1/2}$. The curves adopt $R_{1/2}=f_{\mathrm{size}}\,R_{200}$, with $f_{\mathrm{size}}$ drawn from a lognormal distribution of median $0.015$ and dispersion $\sigma_{\log f}=0.1$~dex. The points are the ratios $R_{\rm TDE}/R_{1/2}$ from Table~\ref{tab:hosts}, and the dashed curves are again the best-fit Gaussians in logarithmic radius.

The observed radii agree with the simulation prediction to leading order, given how small the sample is. The events do cover a wide range of distances, and they were found by a wide mix of methods, surveys, and instruments. The clearest disagreement is that \texttt{ROMULUS} places $\sim30\%$ of WBHs at $10\lesssim R_{\bullet}/R_{1/2}\lesssim100$, where no source is known. For the half-light radii of these six hosts, that range is an angular separation of $\sim20$--$1200$~arcsec between the TDE and the center of the parent. An off-nuclear TDE that far from its parent would look hostless in a transient search, and would be unlikely to pass a TDE filter, so the absence may be a selection effect. A further useful property of TDEs is that they can reveal WBHs much closer to the nucleus, even when no stellar counterpart is visible in pre-TDE, ground-based imaging. They therefore reach a population that dynamical searches cannot. In Fig.~\ref{fig:R_scp}, no dynamical WBH lies at $R_{\bullet,\mathrm{p}}/R_{1/2}<0.5$: a dynamical inference depends on detecting and modeling the stellar counterpart, which is unlikely, or at least much harder, so deep in the stellar halo of the parent, as we show below.

\subsection{Stellar Counterparts}
\label{sec:stellar_counterparts}

\begin{figure*}[t]
\centering
\includegraphics[width=0.8\textwidth]{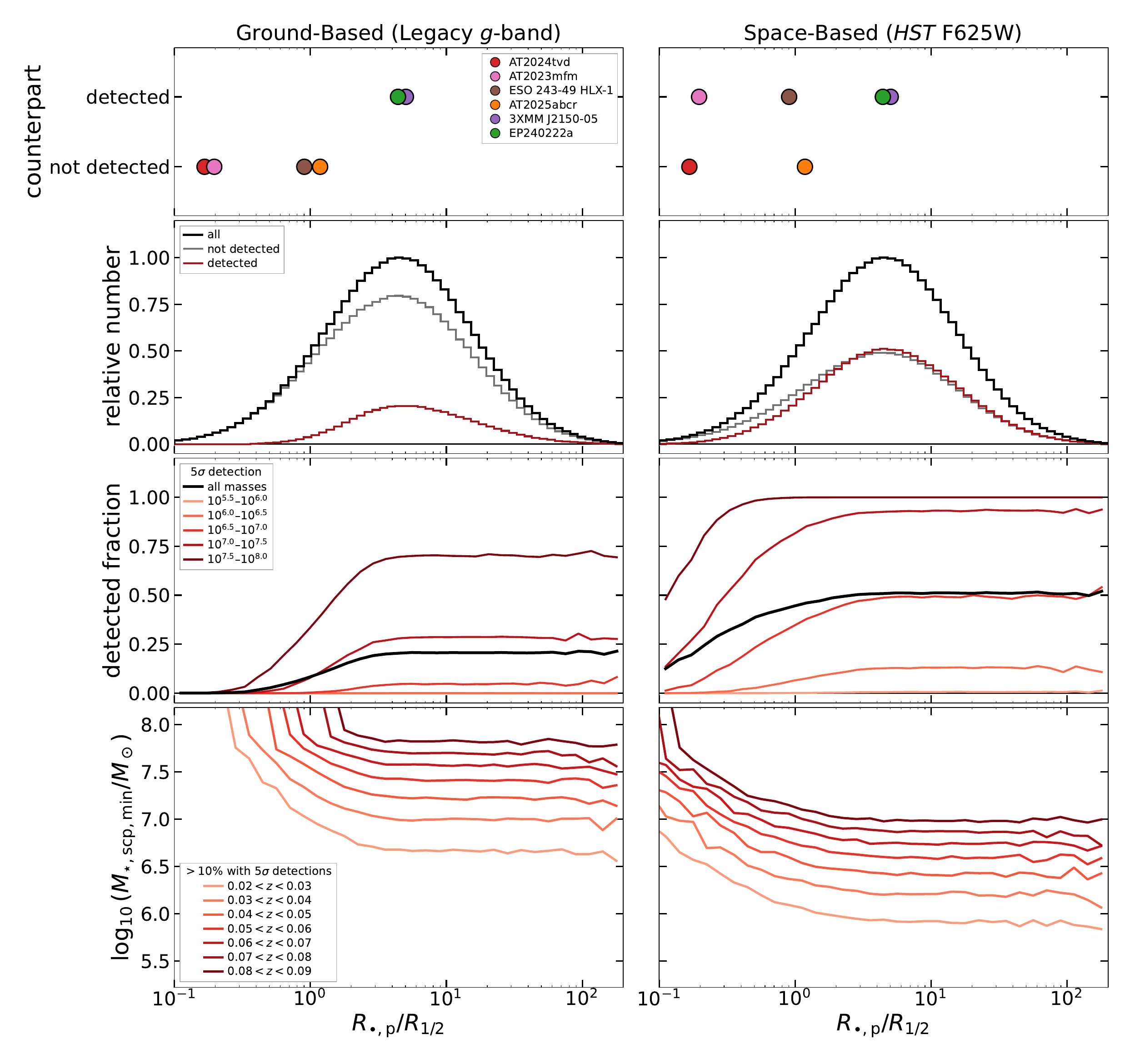}
\caption{Detectability of a stellar counterpart around a WBH.
The top row shows each off-nuclear TDE at its measured $R_{\bullet,\mathrm{p}}/R_{1/2}$ and whether the stellar counterpart is detected (see \S\ref{sec:sample} and Table~\ref{tab:hosts}).
In every row, the left column is Legacy Surveys $g$-band imaging, and the right column is a space-based \hst-like exposure.
The remaining rows are the results of the simulation described in \S\ref{sec:stellar_counterparts}.
The second row is the radial distribution of the simulated stellar counterparts: black is the full input population, gray the systems below $5\sigma$, and red those detected at $\geq5\sigma$.
The third row is the detected fraction in bins of $M_{\star,\mathrm{scp}}$ as a function of $R_{\bullet,\mathrm{p}}/R_{1/2}$.
The bottom row is the lowest $M_{\star,\mathrm{scp}}$ at which at least 10\% of systems at that radius and redshift are detected at $5\sigma$.
From the ground, no stellar counterpart can be detected in the inner halo ($R_{\bullet,\mathrm{p}}/R_{1/2}<1$); even from space, only a fraction is detected there, and only for the combination of higher $M_{\star,\mathrm{scp}}$ and lower $z$.}
\label{fig:stellar_count}
\end{figure*}

In ground-based, pre-TDE imaging, at the depth and seeing of the Legacy Surveys, a stellar counterpart is detected only for the two most offset events, 3XMM~J2150$-$05 and EP240222a, both at $R_{\bullet,\mathrm{p}}/R_{1/2}\gg1$. Both counterparts are consistent with a UCD or a massive stellar cluster \citep{Jin2025,Lin2018,Lin2020}. For the other sources, those with $R_{\bullet,\mathrm{p}}/R_{1/2}\lesssim1$, the pre-TDE ground-based images show no stellar overdensity at the transient position (Fig.~\ref{fig:1} and the top-left panel of Fig.~\ref{fig:stellar_count}).

Even space-based images, deeper and at higher resolution, recover a counterpart for only part of the sample. AT2024tvd and AT2025abcr were observed shortly after the TDE, and the emission at the transient position is still dominated by the TDE (see \S\ref{sec:sample}), with no stellar component needed to explain any of it. A bound stellar population is nevertheless required in these systems (see \S\ref{sec:intro}).

In this section we test whether these observed trends in the detectability of stellar counterparts can be understood, and recovered, if we adopt the \texttt{ROMULUS} results (\S\ref{sec:romulus}): (i) WBHs follow the radial distribution in Fig.~\ref{fig:R_dist}, and (ii) most have been stripped of the stellar envelope (disk and/or halo) of their original galaxy, but can still retain a compact stellar cluster.

To do so, we perform the following simulation:
\begin{itemize}
\item We generate $10^{4}$ halos with $12.5\leq\log_{10}(M_{200}/M_\odot)<13.5$, the halo-mass range of our sample in Table~\ref{tab:hosts}. Each halo is assigned an $M_{\star}$ from the SHMR (Eq.~\ref{eq:SHMR}), and the hosts follow the GSMF (Eq.~\ref{eq:gsmf}). Redshifts are drawn uniformly from $0.02<z<0.09$, the range spanned by our events.
\item Each host is assumed to be an early-type galaxy, with half-light radius $R_{1/2}=f_{\mathrm{size}}R_{200}$. The factor $f_{\mathrm{size}}$ is lognormally distributed, with median $0.015$ and dispersion $0.1$~dex. The light profile is a de Vaucouleurs bulge plus an exponential envelope, and the envelope is assigned $0.60\pm0.2$ of the light, drawn from a Gaussian, as is typical of early-type galaxies \citep{Huang2013}.
\item The number of WBHs in a halo is a Poisson draw from Eq.~\ref{eq:Nwbh}. Their halo-centric radii follow the Gaussian fits in Fig.~\ref{fig:R_dist}, and the sky projection is isotropic.
\item Each WBH carries a bound stellar counterpart with a King profile. $M_{\star,\mathrm{scp}}$ is drawn log-uniformly in $\log M_{\star,\mathrm{scp}}$ between $10^{5.5}$ and $10^{8}\,M_\odot$, the range spanned by our sample, and the half-light radius is drawn uniformly in $\log R_{\mathrm{e}}$ between $10$ and $100$~pc. We assume an old ($10$~Gyr) stellar population, consistent with the stellar counterparts of 3XMM~J2150$-$05 and EP240222a.
\item Every system is imaged twice: once in the Legacy Surveys $g$ band, with $1\arcsec$ seeing and a $5\sigma$ point-source depth of $24.7$~mag, and once in a $600$~s \hst\ WFC3 F625W exposure, with $0\farcs1$ resolution and a depth of $26.7$~mag. Noise is added as the photon noise of the sky and of the galaxy, set so that an isolated point source at the quoted depth is a $5\sigma$ detection.
\item For each image we subtract a model of the parent-galaxy light and perform aperture photometry on the residual at the position of the stellar counterpart. We record whether the counterpart is recovered at $\gtrsim 5\sigma$ (examples of the resulting residual images are shown in Appendix~\ref{app:c}, Fig.~\ref{fig:cp_examples}).
\end{itemize}

A summary of the results of the simulation is shown in Fig.~\ref{fig:stellar_count}, in which the left panels show the ground-based results and the right panels the space-based results.

In the panels immediately below the observations, we show the radial distribution of all WBHs in the $10^{4}$ halos. In black is the input distribution, the combination of the two \texttt{ROMULUS} radial profiles in Fig.~\ref{fig:R_dist}. In gray is the distribution of those whose stellar counterpart is not detected, and in red the distribution of those whose stellar counterpart is detected. From the ground, only a small fraction ($\sim20\%$) of the simulated population is detected, and effectively none at $R_{\bullet,\mathrm{p}}/R_{1/2}<1$, so the detected distribution is shifted to larger radii relative to the input population. Even from space, only $\sim50\%$ of the simulated population is detected.

The middle and bottom panels show how this detectability scales with radius, with $M_{\star,\mathrm{scp}}$, and with $z$. In both the ground-based and the space-based images, the detected fraction in a given mass range is flat in the outer halo. There, the detectability of the counterpart is set by the combination of $z$ and $M_{\star,\mathrm{scp}}$ for an isolated source, at the depth of the image. As the WBH enters the stellar halo of the parent, $R_{\bullet,\mathrm{p}}/R_{1/2}\lesssim$ a few, the fraction decreases in both the space-based and the ground-based simulations. Within one half-light radius, $R_{\bullet,\mathrm{p}}/R_{1/2}<1$, the decline is stronger. At every $z$, and for any $M_{\star,\mathrm{scp}}<10^{7.5}\,M_{\odot}$, effectively no counterpart is detected in the ground-based images, and even a space-based observation leaves a large fraction undetected.

In the bottom panels, we instead show the minimum mass, $M_{\star,\mathrm{scp,min}}$, as a function of radius. This is the mass at which at least 10\% of the simulated population in that mass and $z$ bin is detected. The pattern is the same. At $R_{\bullet,\mathrm{p}}/R_{1/2}\gtrsim$ a few, the curves are flat, and the minimum mass is set by the depth of the image for an isolated source. The minimum mass increases sharply as the position moves into the inner stellar halo of the parent.

The simulation shows that if the radial distribution of WBHs follows \texttt{ROMULUS}, the observed pattern follows as well. Ground-based, pre-TDE images detect a counterpart only at large separation, and space-based imaging still misses a substantial fraction of the counterparts, even under the assumption that every WBH retains a stellar counterpart. The limit is the surface brightness of the parent galaxy. For a given PSF and image depth, the parent outshines the stellar counterpart, and it does so more strongly deeper within the stellar halo.

These results are conservative: two further effects would make the stellar counterpart even harder to detect. First, the simulation assumes no radial dependence of $M_{\star,\mathrm{scp}}$. If the average surviving mass decreases toward smaller radii, because those systems have been stripped for longer, the suppression of detectability at low radii would be even stronger. Second, the post-TDE space-based images are taken to be entirely stellar. The calculation leaves out the TDE light, which outshines the stellar emission and does so most strongly at early times, as is the case for AT2024tvd and AT2025abcr. Including that light would lower the detected fraction at all radii and would make the fraction depend on the time of the observation after the TDE. Both effects work in the same direction, making the counterpart even harder to detect.

The same surface-brightness limit accounts for the absence of dynamically detected WBHs within $0.5$ half-light radii of the parent (Fig.~\ref{fig:R_scp}). Inside that radius the parent hides the stellar counterpart, and a dynamical measurement requires its detection and kinematic modeling at extremely high signal-to-noise. Off-nuclear TDEs can be used to discover this inner population through their bright transient emission, even when no stellar counterpart can be detected. This is a particularly important population, as it is the closest to forming an extreme-mass-ratio binary with the central black hole, and therefore a likely progenitor population of LISA sources \citep{AmaroSeoane2017}.

\section{Conclusion}\label{sec:conclusion}

The growing sample of secure off-nuclear TDEs provides direct empirical evidence for, and a path to the study of, the long-predicted population of off-nuclear MBHs ($M_{\bullet}\geq 10^{4}\,M_{\odot}$). Despite the small sample size, these events already show a set of interesting demographic patterns:

\begin{itemize}
    \item All occur in massive, early-type galaxies with $M_{\star}\simeq 10^{11}\,M_{\odot}$, a range that very rarely hosts nuclear TDEs \citep{Hammerstein_21,Yao2023}, but the same galaxies in which dynamically inferred off-nuclear black holes are known \citep{Seth2014,Voggel2019,Taylor2025}.
    \item By counting companions with $M_{\star}\ge 10^{9}\,M_{\odot}$ in their halos, we show that these off-nuclear TDEs, like the dynamically detected WBHs, mostly inhabit galaxy groups and clusters, while nuclear TDEs are mostly in isolated galaxies.
    \item Their deprojected radii span nearly two orders of magnitude, from $R_{\bullet}/R_{200}\simeq 0.003$ to $\simeq 0.1$, or from $R_{\bullet,\mathrm{p}}/R_{1/2}\simeq 0.2$ to $\simeq 6$.
     \item From the ground, a stellar counterpart to the WBH is detected only when it lies in the outskirts of the parent galaxy ($R_{\bullet,\mathrm{p}}/R_{1/2}\gg 1$). Closer in, even space-based imaging recovers a counterpart for only part of the sample.
\end{itemize}

In this Letter, we showed that these trends follow from hierarchical assembly, as realized in the \texttt{ROMULUS} cosmological simulation, if these WBHs are formed via tidal stripping in minor mergers. Combining the WBH occupation in the simulation with the local GSMF, we construct the volumetric density of WBHs, $\phi_{\rm WBH}(M_{\star})$, which, as we show, peaks at $\log_{10}(M_{\star}/M_{\odot})\sim 11$ and places the central 68\% of local WBHs in $10.65\lesssim\log_{10}(M_{\star}/M_{\odot})\lesssim 11.30$, naturally explaining why the entire off-nuclear sample lies in galaxies in this mass range. Over the same interval, early-type galaxies outnumber late-type galaxies, and the number of companion galaxies increases with halo mass, so the red colors and the denser environments follow from the same parent-galaxy preference. We also show that if the observed off-nuclear TDE rate as a function of parent-galaxy mass is, to leading order, proportional to the volumetric density of WBHs, $\rho^{\rm off-nuclear}(M_{\star})\propto\phi_{\rm WBH}(M_{\star})$, then in high-mass galaxies, $\log_{10}(M_{\star}/M_{\odot})\gg 10.7$, the off-nuclear rate is at least comparable to, and likely higher than, the nuclear rate. This result is compatible with the current rate constraints from ZTF \citep{Stein2026}.

The radial distribution is, to leading order, a Gaussian in $\log_{10}(R_{\bullet}/R_{200})$, and the six events lie on it. The strongest disagreement is that \texttt{ROMULUS} places $\sim30\%$ of WBHs at $10\lesssim R_{\bullet}/R_{1/2}\lesssim100$, where no off-nuclear TDE is known. At the redshifts of this sample, that separation is $\sim20$--$1200$~arcsec, far enough that the flare would appear hostless and would be unlikely to be identified as an off-nuclear TDE candidate for follow-up. This population should nevertheless exist.

We also show that if the radial distribution of WBHs follows \texttt{ROMULUS}, the contrast between the stellar-counterpart light and the surface brightness of the stellar halo necessarily produces the observed pattern of counterpart detections. Ground-based imaging detects a counterpart only outside a few half-light radii, and even space-based imaging, at these redshifts and $M_{\star,\mathrm{scp}}$, cannot detect every system.

This effect explains why no dynamically detected WBH lies at $R_{\bullet,\mathrm{p}}/R_{1/2}<0.5$: inside that radius the light of the parent hides the stellar counterpart, and a dynamical measurement cannot be made without it. Off-nuclear TDEs are then the only way to reach this inner-halo WBH population, which is also the one closest to forming an extreme-mass-ratio binary with the central black hole, and these binaries are relevant for LISA \citep{AmaroSeoane2017}.

Taken together, these results support a simple interpretation: off-nuclear TDEs are the accreting counterparts of the WBHs long predicted by cosmological simulations, and those black holes are mostly formed in minor mergers. As the sample grows, parent galaxies should continue to be drawn from high-mass early-type galaxies and should be easily found in groups and clusters, which could be used as a watch list for off-nuclear TDE searches in surveys such as the Legacy Survey of Space and Time \citep[LSST,][]{Ivezic2019}.
\\

\textit{Acknowledgments} --
MG is grateful to Suvi Gezari and Andrew Mummery for their comments on an early version of this paper, and to Angelo Ricarte and Barak Rom for valuable discussions on WBHs and off-nuclear TDEs. MG is grateful for support from NASA and the Space Telescope Science Institute.

\appendix

\section{From $M_{\star}$ and $R_{1/2}$ to $M_{200}$ and $R_{200}$}
\label{app:1}

We estimate the virial mass ($M_{200}$) and corresponding virial radius ($R_{200}$)
of each galaxy from its $M_{\star}$, projected half-light radius
($R_{1/2}$), and redshift ($z$) using a Bayesian framework that combines the
SHMR with the empirical size--halo relation. The
SHMR provides a monotonic mapping between stellar and halo mass, reflecting the
efficiency of baryon conversion into stars across cosmic time. We adopt the
functional form of \citet{Moster2013}, as shown in Equation~\ref{eq:SHMR}, with an
intrinsic scatter of $\sigma_{\log M_*} = 0.1~\mathrm{dex}$.

Galaxy size information is incorporated through the empirical scaling between the
half-light radius and halo radius, $R_{1/2} = f_{\mathrm{size}}\, R_{200}$, which
encapsulates the structural coupling between galaxies and their host halos.

For early-type (quenched) systems, we adopt a prior $f_{\mathrm{size}} = 0.015$
with a lognormal dispersion $\sigma_{\log f} = 0.1~\mathrm{dex}$, consistent with
observed size--halo relations for massive ellipticals \citep[e.g.,][]{Kravtsov2013}.
Both $M_{200}$ and $f_{\mathrm{size}}$ are treated as free parameters, and their
joint posterior distribution is sampled using \texttt{emcee}
\citep{Foreman-Mackey_13}.

The likelihood simultaneously constrains the stellar and structural observables as
\begin{multline}
\ln \mathcal{L} = -\frac{1}{2} \left[
\frac{(\log M_*^{\mathrm{obs}} - \log M_*^{\mathrm{model}}(M_{200}))^2}{\sigma_{\log M_*}^2} \right] + \\
\frac{1}{2} \left[ \frac{(\log R_{1/2}^{\mathrm{obs}} - \log(f_{\mathrm{size}} R_{200}(M_{200})))^2}{\sigma_{\log R}^2}
\right],
\end{multline}
where $\sigma_{\log R} = 0.1~\mathrm{dex}$ represents the intrinsic scatter in the
size--halo relation.

For each Markov chain Monte Carlo (MCMC) sample, the virial radius is computed from the corresponding halo
mass using
\begin{equation}
M_{200} = \frac{4}{3}\pi R_{200}^3\, 200\, \rho_{\mathrm{c}}(z),
\end{equation}
where $\rho_{\mathrm{c}}(z)$ is the critical density of the Universe at redshift
$z$. The critical density evolves with redshift as
$\rho_{\mathrm{c}}(z) = \frac{3 H^2(z)}{8\pi G}$,
where $G$ is the gravitational constant and $H(z)$ is the Hubble parameter at
redshift $z$. The latter is obtained from the cosmological expansion law in a flat
$\Lambda$CDM model,
$H(z) = H_0 \sqrt{\Omega_{\mathrm{m}} (1+z)^3 + \Omega_{\Lambda}}$,
so that $\rho_{\mathrm{c}}(z)$ increases with redshift. This evolving critical
density defines the overdensity threshold used to calculate virial masses and
radii.

Overall, this approach combines $M_{\star}$ and galaxy size information to infer
self-consistent halo properties while naturally incorporating observational
scatter and empirical scaling uncertainties.

\section{Galaxy Masses}
\label{app:2}

We estimate parent-galaxy $M_{\star}$ in a homogeneous manner using optical
photometry from the Legacy Surveys Data Release~10 \citep{Dey2019}, following the color--dependent
mass-to-light ratio prescription of \citet{Baldry2012}. Total galaxy fluxes are
measured from the best-fitting \textsc{Tractor} S\'ersic models in the \emph{g}
and \emph{i} bands.

All photometry is corrected for Galactic extinction using the dust maps of
\citet{Schlafly_2011} and the extinction law of \citet{Cardelli1989}. Stellar
masses are then computed from the extinction-corrected \emph{i}-band luminosity
and the rest-frame color $(g-i)_0$ via
\begin{equation}
\log_{10}(M_\star/L_i) = a + b\,(g-i)_0,
\end{equation}
with coefficients $a=-0.68$ and $b=0.73$, as calibrated by
\citet{Baldry2012}. 

The uncertainty on each $M_{\star}$ is the \citet{Taylor2011} mass-to-light term of $0.1\,\mathrm{dex}$, the same value for every galaxy, added in quadrature to that galaxy's own uncertainty on $(g-i)_0$ and on the $i$-band luminosity,
$\sigma_{\log M}=\sqrt{0.1^{2}+(0.73\,\sigma_{g-i})^{2}+\sigma_{\log L_i}^{2}}$.
We assume no internal extinction within the parent galaxies,
which is a reasonable approximation for quenched systems.

By using S\'ersic luminosities, as in \citet{Baldry2012}, we ensure that the
effective aperture is not fixed but instead scales with galaxy distance and
physical size. This guarantees that $M_{\star}$ values are measured consistently
across the sample and can be meaningfully compared. Importantly, by adopting the
exact same methodology as \citet{Baldry2012}, we ensure that our inferred values of $M_{\star}$ are directly comparable to those entering the GSMF
used throughout this work.

Where a galaxy has no $i$-band detection, $g$ and $r$ are corrected for
Galactic extinction, then shifted to the rest frame with the
\citet{Chilingarian2012} $K$-corrections in redshift and $g-r$,
and $(g-i)_0$ is estimated by inverting
\begin{equation}
(r-i) = -0.005621 + 0.458938\,(g-i) - 0.099596\,(g-i)^{2},
\end{equation}
the quadratic fit to SDSS DR18 model colors (rms $0.04\,\mathrm{mag}$ in $r-i$),
applied to that rest-frame $g-r$. The rest-frame $i$ magnitude is
$i = g_K - (g-i)_0$. The
$0.04\,\mathrm{mag}$ scatter is included in $\sigma_{g-i}$, and the
mass-to-light relation above is unchanged. The solar absolute magnitude is $M_{i,\odot}=4.53$ in the
AB system.

\section{Additional Figures and Tables}
\label{app:c}

The properties of the dynamically detected WBHs are listed in Table~\ref{tab:dyn}. Tables~\ref{tab:nuc_ztf} and \ref{tab:nuc_ero} list the properties of the nuclear TDE hosts defined in \S\ref{sec:comp_samp}. 

\begin{deluxetable*}{lcc|lccccc|c}
\tablecaption{Dynamically detected WBHs \label{tab:dyn}}
\tabletypesize{\scriptsize}
\tablehead{
\multicolumn{3}{c|}{Source} & \multicolumn{6}{c|}{Parent Galaxy} & \colhead{Stellar Counterpart} \\
\colhead{Name} & \colhead{$R_{\bullet,{\rm p}}$} & \multicolumn{1}{c|}{$R_{\bullet}$} &
\colhead{Name} & \colhead{Distance} & \colhead{$\log_{10} M_\star$} & \colhead{$R_{1/2}$} & \colhead{$\log_{10} M_{200}$} & \multicolumn{1}{c|}{$R_{200}$} &
\colhead{$M_{\star,{\rm scp}}$} \\
\colhead{} & \colhead{(arcsec, kpc)} & \multicolumn{1}{c|}{(kpc)} &
\colhead{} & \colhead{(Mpc)} & \colhead{($M_\odot$)} & \colhead{(arcsec, kpc)} & \colhead{($M_\odot$)} & \multicolumn{1}{c|}{(kpc)} &
\colhead{($M_\odot$)}
}
\startdata
G1 & 10581, 39.5 & $45.6_{-5.6}^{+27.4}$ & M31 & 0.77 & $11.00\pm0.10$ & 2384, 8.9 & $13.0\pm0.2$ & $460_{-70}^{+90}$ & $10^{6.32\pm0.10}$ \\
B023-G078 & 1179, 4.4 & $5.1_{-0.6}^{+3.1}$ & M31 & 0.77 & $11.00\pm0.10$ & 2384, 8.9 & $13.0\pm0.2$ & $460_{-70}^{+90}$ & $10^{6.55\pm0.10}$ \\
M59-UCD3 & 129, 10.3 & $11.9_{-1.5}^{+7.2}$ & M59 & 16.5 & $10.75\pm0.10$ & 41.5, 3.3 & $12.4\pm0.2$ & $280_{-40}^{+50}$ & $10^{7.91\pm0.10}$ \\
M59cO & 126, 10.1 & $11.6_{-1.4}^{+7.0}$ & M59 & 16.5 & $10.75\pm0.10$ & 41.5, 3.3 & $12.4\pm0.2$ & $280_{-40}^{+50}$ & $10^{7.61\pm0.10}$ \\
UCD736 & 724, 57.9 & $66.8_{-8.2}^{+39.9}$ & M59 & 16.5 & $10.75\pm0.10$ & 41.5, 3.3 & $12.4\pm0.2$ & $280_{-40}^{+50}$ & $10^{7.19\pm0.10}$ \\
M60-UCD1 & 88, 7.0 & $8.1_{-1.0}^{+4.8}$ & M60 & 16.5 & $11.20\pm0.10$ & 55.6, 4.4 & $13.4\pm0.2$ & $590_{-100}^{+110}$ & $10^{7.86\pm0.10}$ \\
VUCD3 & 179, 14.3 & $16.6_{-2.1}^{+10.0}$ & M87 & 16.5 & $11.18\pm0.10$ & 55.5, 4.4 & $13.3\pm0.2$ & $570_{-90}^{+110}$ & $10^{7.24\pm0.10}$ \\
Fornax UCD3 & 500, 50.6 & $58.6_{-7.3}^{+34.8}$ & NGC 1399 & 20.9 & $11.21\pm0.10$ & 59.2, 6.0 & $13.5\pm0.2$ & $630_{-100}^{+130}$ & $10^{7.26\pm0.10}$ \\
$\omega$~Cen & --- & $6.4$ & Milky Way & --- & $10.42\pm0.10$ & ---, $5.8\pm0.4$ & $11.9\pm0.1$ & $190_{-20}^{+20}$ & $10^{6.524\pm0.005}$ \\
\enddata
\tablecomments{$R_{\bullet}$ is deprojected for an isotropic orientation, as in Table~\ref{tab:hosts}, except for $\omega$~Cen, where the value is the measured galactocentric radius and no sky-projected offset is defined. Distances are redshift-independent: 16.5~Mpc for M60, M59, and M87 \citep{Mei2007}, 20.9~Mpc for NGC~1399 \citep{Blakeslee2009}, and 0.77~Mpc for M31 \citep{Karachentsev2004}, as adopted by \citet{Pechetti2022}. The Milky Way has no external distance. Projected offsets were converted to kpc with these distances; a Hubble-flow redshift is not a distance in Virgo, Fornax, or the Local Group. $M_{\star,\mathrm{scp}}$ is the $M_{\star}$ of the compact system; for $\omega$~Cen it is the cluster mass of \citet{Baumgardt2018}. The Milky Way half-light radius is the value from \citet{Lian2024}.}
\end{deluxetable*}

\startlongtable
\begin{deluxetable*}{lccccc}
\tablecaption{Nuclear TDE hosts from ZTF\label{tab:nuc_ztf}}
\tabletypesize{\scriptsize}
\tablehead{
\colhead{Name} & \colhead{$z$} & \colhead{$\log_{10}(M_\star/M_\odot)$} &
\colhead{$\log_{10}(M_{200}/M_\odot)$} & \colhead{$R_{200}$} & \colhead{$N$} \\
\colhead{} & \colhead{} & \colhead{} & \colhead{} & \colhead{(kpc)} & \colhead{}
}
\startdata
AT2018hyz & 0.04573 & $9.66\pm0.10$ & $11.3\pm0.1$ & $120_{-10}^{+10}$ & $<0.01$ \\
AT2019azh & 0.022 & $9.88\pm0.10$ & $11.4\pm0.1$ & $130_{-10}^{+10}$ & $<0.01$ \\
AT2019bhf & 0.1206 & $10.27\pm0.11$ & $11.7\pm0.1$ & $160_{-10}^{+10}$ & $0.08_{-0.08}^{+0.29}$ \\
AT2019ehz & 0.074 & $9.71\pm0.11$ & $11.3\pm0.1$ & $120_{-10}^{+10}$ & $1.09_{-0.09}^{+0.29}$ \\
AT2019lwu & 0.117 & $9.70\pm0.10$ & $11.3\pm0.1$ & $120_{-10}^{+10}$ & $<0.01$ \\
AT2019qiz & 0.0151 & $10.19\pm0.10$ & $11.6\pm0.1$ & $150_{-10}^{+10}$ & $1.02_{-0.02}^{+0.14}$ \\
AT2019vcb & 0.089 & $9.68\pm0.11$ & $11.3\pm0.1$ & $120_{-10}^{+10}$ & $<0.01$ \\
AT2020mbq & 0.093 & $9.48\pm0.11$ & $11.2\pm0.1$ & $110_{-10}^{+10}$ & $0.01_{-0.01}^{+0.10}$ \\
AT2020neh & 0.062 & $9.61\pm0.10$ & $11.3\pm0.1$ & $120_{-10}^{+10}$ & $<0.01$ \\
AT2020nov & 0.084 & $10.60\pm0.10$ & $12.1\pm0.2$ & $220_{-20}^{+40}$ & $0.11_{-0.11}^{+0.33}$ \\
AT2020opy & 0.159 & $10.19\pm0.11$ & $11.6\pm0.1$ & $150_{-10}^{+10}$ & $<0.01$ \\
AT2020pj & 0.068 & $9.97\pm0.11$ & $11.5\pm0.1$ & $140_{-10}^{+10}$ & $<0.01$ \\
AT2020vdq & 0.044 & $9.20\pm0.11$ & $11.1\pm0.1$ & $100_{-10}^{+10}$ & $0.02_{-0.02}^{+0.13}$ \\
AT2020vwl & 0.0325 & $9.91\pm0.11$ & $11.4\pm0.1$ & $130_{-10}^{+10}$ & $<0.01$ \\
AT2020wey & 0.02738 & $9.66\pm0.11$ & $11.3\pm0.1$ & $120_{-10}^{+10}$ & $<0.01$ \\
AT2020zso & 0.061 & $10.08\pm0.10$ & $11.6\pm0.1$ & $140_{-10}^{+10}$ & $<0.01$ \\
AT2021axu & 0.192 & $10.21\pm0.10$ & $11.6\pm0.1$ & $150_{-10}^{+10}$ & $0.09_{-0.09}^{+0.29}$ \\
AT2021jjm & 0.153 & $9.32\pm0.10$ & $11.1\pm0.1$ & $100_{-10}^{+10}$ & $<0.01$ \\
AT2021mhg & 0.073 & $9.61\pm0.11$ & $11.3\pm0.1$ & $120_{-10}^{+10}$ & $1$ \\
AT2021nwa & 0.047 & $10.14\pm0.11$ & $11.6\pm0.1$ & $150_{-10}^{+10}$ & $<0.01$ \\
AT2021qth & 0.0805 & $9.32\pm0.10$ & $11.1\pm0.1$ & $100_{-10}^{+10}$ & $0.10_{-0.10}^{+0.32}$ \\
AT2021utq & 0.127 & $9.78\pm0.11$ & $11.4\pm0.1$ & $120_{-10}^{+10}$ & $0.11_{-0.11}^{+0.31}$ \\
AT2022aee & 0.124 & $10.08\pm0.11$ & $11.6\pm0.1$ & $140_{-10}^{+10}$ & $<0.01$ \\
AT2022bdw & 0.03782 & $10.32\pm0.10$ & $11.7\pm0.1$ & $170_{-10}^{+10}$ & $1$ \\
AT2022dyt & 0.072 & $10.51\pm0.10$ & $12.0_{-0.1}^{+0.2}$ & $200_{-20}^{+30}$ & $<0.01$ \\
AT2022exr & 0.096 & $10.38\pm0.11$ & $11.8\pm0.1$ & $170_{-10}^{+20}$ & $0.24_{-0.24}^{+0.45}$ \\
AT2022hvp & 0.120 & $10.62\pm0.11$ & $12.1\pm0.2$ & $220_{-30}^{+40}$ & $0.17_{-0.17}^{+0.40}$ \\
AT2022lri & 0.03275 & $9.85\pm0.10$ & $11.4\pm0.1$ & $130_{-10}^{+10}$ & $<0.01$ \\
AT2022pna & 0.095 & $9.88\pm0.10$ & $11.4\pm0.1$ & $130_{-10}^{+10}$ & $<0.01$ \\
AT2022upj & 0.054 & $10.40\pm0.10$ & $11.8\pm0.1$ & $180_{-10}^{+20}$ & $0.02_{-0.02}^{+0.13}$ \\
AT2022wtn & 0.049 & $10.25\pm0.11$ & $11.7\pm0.1$ & $160_{-10}^{+10}$ & $<0.01$ \\
AT2023adr & 0.132 & $10.11\pm0.11$ & $11.6\pm0.1$ & $140_{-10}^{+10}$ & $<0.01$ \\
AT2023mhs & 0.0482 & $9.94\pm0.10$ & $11.5\pm0.1$ & $130_{-10}^{+10}$ & $<0.01$ \\
AT2023wdb & 0.060 & $10.25\pm0.11$ & $11.7\pm0.1$ & $160_{-10}^{+10}$ & $0.26_{-0.26}^{+0.49}$ \\
AT2024aepd & 0.0835 & $10.06\pm0.11$ & $11.5\pm0.1$ & $140_{-10}^{+10}$ & $<0.01$ \\
AT2024as & 0.130 & $10.11\pm0.10$ & $11.6\pm0.1$ & $140_{-10}^{+10}$ & $0.02_{-0.02}^{+0.14}$ \\
AT2024bgz & 0.0585 & $10.51\pm0.10$ & $12.0_{-0.1}^{+0.2}$ & $200_{-20}^{+30}$ & $0.01_{-0.01}^{+0.12}$ \\
AT2024gre & 0.120 & $9.88\pm0.11$ & $11.4\pm0.1$ & $130_{-10}^{+10}$ & $<0.01$ \\
AT2024gxr & 0.051 & $10.44\pm0.11$ & $11.9_{-0.1}^{+0.2}$ & $180_{-20}^{+20}$ & $0.01_{-0.01}^{+0.11}$ \\
AT2024pvu & 0.048 & $10.36\pm0.11$ & $11.8\pm0.1$ & $170_{-10}^{+20}$ & $0.14_{-0.14}^{+0.34}$ \\
AT2024qab & 0.1398 & $10.22\pm0.11$ & $11.7\pm0.1$ & $150_{-10}^{+10}$ & $0.04_{-0.04}^{+0.20}$ \\
AT2024ule & 0.111 & $10.15\pm0.10$ & $11.6\pm0.1$ & $150_{-10}^{+10}$ & $0.23_{-0.23}^{+0.46}$ \\
AT2024wsd & 0.063 & $9.54\pm0.10$ & $11.2\pm0.1$ & $110_{-10}^{+10}$ & $<0.01$ \\
AT2024ymu & 0.120 & $10.90\pm0.10$ & $12.8_{-0.2}^{+0.3}$ & $360_{-60}^{+80}$ & $<0.01$ \\
AT2024yqo & 0.06168 & $10.72\pm0.10$ & $12.3\pm0.2$ & $260_{-40}^{+50}$ & $0.03_{-0.03}^{+0.16}$ \\
\enddata
\tablecomments{Nuclear TDE hosts from ZTF, as defined in \S\ref{sec:comp_samp}. $M_\star$ is from Appendix~\ref{app:2}, $M_{200}$ and $R_{200}$ from Appendix~\ref{app:1}, and $N$ from Appendix~\ref{app:3}. Counts below $0.01$ are written as upper limits.}
\end{deluxetable*}

\startlongtable
\begin{deluxetable*}{lccccc}
\tablecaption{Nuclear TDE hosts from eROSITA\label{tab:nuc_ero}}
\tabletypesize{\scriptsize}
\tablehead{
\colhead{Name} & \colhead{$z$} & \colhead{$\log_{10}(M_\star/M_\odot)$} &
\colhead{$\log_{10}(M_{200}/M_\odot)$} & \colhead{$R_{200}$} & \colhead{$N$} \\
\colhead{} & \colhead{} & \colhead{} & \colhead{} & \colhead{(kpc)} & \colhead{}
}
\startdata
eRASSt J0114$-$59 & 0.160 & $11.02\pm0.10$ & $13.1\pm0.3$ & $450_{-80}^{+100}$ & $1.18_{-1.01}^{+1.02}$ \\
eRASSt J0132+12 & 0.132 & $10.35\pm0.11$ & $11.8\pm0.1$ & $170_{-10}^{+20}$ & $0.13_{-0.13}^{+0.35}$ \\
eRASSt J0141$-$44 & 0.090 & $9.80\pm0.10$ & $11.4\pm0.1$ & $120_{-10}^{+10}$ & $<0.01$ \\
eRASSt J0439$-$65 & 0.152 & $11.06\pm0.08$ & $13.2\pm0.2$ & $480_{-70}^{+80}$ & $2.87\pm1.54$ \\
eRASSt J0608$-$43 & 0.080 & $10.24\pm0.10$ & $11.7\pm0.1$ & $160_{-10}^{+10}$ & $0.03_{-0.03}^{+0.15}$ \\
eRASSt J0634$-$71 & 0.130 & $10.11\pm0.10$ & $11.6\pm0.1$ & $140_{-10}^{+10}$ & $0.01_{-0.01}^{+0.12}$ \\
eRASSt J0644$-$60 & 0.120 & $10.21\pm0.10$ & $11.6\pm0.1$ & $150_{-10}^{+10}$ & $0.07_{-0.07}^{+0.26}$ \\
eRASSt J0713+72 & 0.104 & $9.88\pm0.11$ & $11.4\pm0.1$ & $130_{-10}^{+10}$ & $<0.01$ \\
eRASSt J0744+29 & 0.040 & $9.16\pm0.10$ & $11.1\pm0.1$ & $100_{-10}^{+10}$ & $<0.01$ \\
eRASSt J0820+19 & 0.120 & $10.20\pm0.10$ & $11.6\pm0.1$ & $150_{-10}^{+10}$ & $0.12_{-0.12}^{+0.34}$ \\
eRASSt J0823+04 & 0.030 & $10.35\pm0.10$ & $11.8\pm0.1$ & $170_{-10}^{+20}$ & $5.33_{-0.33}^{+0.53}$ \\
eRASSt J0916+06 & 0.090 & $9.73\pm0.10$ & $11.3\pm0.1$ & $120_{-10}^{+10}$ & $<0.01$ \\
eRASSt J0917+52 & 0.1876 & $9.88\pm0.11$ & $11.4\pm0.1$ & $120_{-10}^{+10}$ & $<0.01$ \\
eRASSt J1330+73 & 0.1505 & $10.30\pm0.11$ & $11.7\pm0.1$ & $160_{-10}^{+10}$ & $0.06_{-0.06}^{+0.24}$ \\
eRASSt J1355+31 & 0.1989 & $10.36\pm0.11$ & $11.8\pm0.1$ & $160_{-10}^{+20}$ & $0.02_{-0.02}^{+0.16}$ \\
eRASSt J1447+67 & 0.125 & $10.68\pm0.11$ & $12.3\pm0.2$ & $240_{-40}^{+50}$ & $0.42_{-0.42}^{+0.63}$ \\
eRASSt J1456$-$28 & 0.090 & $10.85\pm0.11$ & $12.7\pm0.3$ & $330_{-60}^{+80}$ & $1.07_{-0.98}^{+0.97}$ \\
eRASSt J1459$-$26 & 0.090 & $9.84\pm0.10$ & $11.4\pm0.1$ & $130_{-10}^{+10}$ & $0.01_{-0.01}^{+0.12}$ \\
eRASSt J1534$-$09 & 0.020 & $10.14\pm0.10$ & $11.6\pm0.1$ & $150_{-10}^{+10}$ & $<0.01$ \\
eRASSt J1610+33 & 0.1309 & $10.40\pm0.11$ & $11.8\pm0.1$ & $170_{-10}^{+20}$ & $<0.01$ \\
eRASSt J1646$-$69 & 0.020 & $10.41\pm0.10$ & $11.8\pm0.1$ & $180_{-10}^{+20}$ & $<0.01$ \\
eRASSt J1714+08 & 0.036 & $10.49\pm0.10$ & $11.9_{-0.1}^{+0.2}$ & $190_{-20}^{+20}$ & $0.07_{-0.07}^{+0.26}$ \\
eRASSt J1901$-$55 & 0.060 & $10.54\pm0.10$ & $12.0_{-0.1}^{+0.2}$ & $200_{-20}^{+30}$ & $0.12_{-0.12}^{+0.34}$ \\
eRASSt J1921$-$50 & 0.060 & $10.58\pm0.04$ & $12.1\pm0.1$ & $210_{-10}^{+10}$ & $0.32_{-0.32}^{+0.55}$ \\
eRASSt J2108$-$56 & 0.040 & $10.61\pm0.10$ & $12.1\pm0.2$ & $220_{-30}^{+40}$ & $0.02_{-0.02}^{+0.15}$ \\
eRASSt J2344$-$35 & 0.100 & $10.20\pm0.10$ & $11.6\pm0.1$ & $150_{-10}^{+10}$ & $<0.01$ \\
\enddata
\tablecomments{Same as Table~\ref{tab:nuc_ztf}, but for the eROSITA hosts.}
\end{deluxetable*}

\begin{figure}
    \centering
    \includegraphics[width=\linewidth]{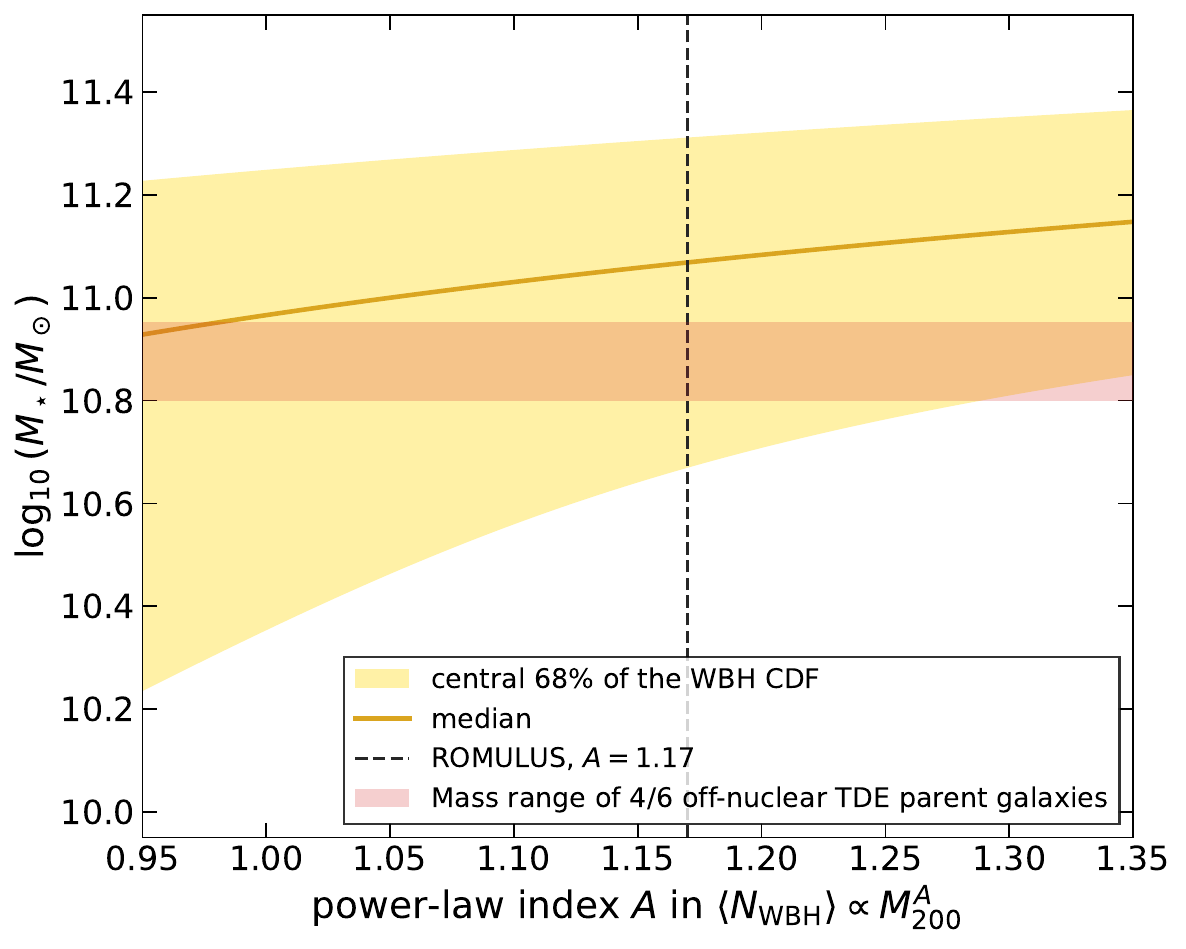}
    \caption{$M_{\star}$ of WBH hosts as a function of the occupation slope $A$ in $\langle N_{\rm WBH}\rangle \propto M_{200}^{A}$. The curve is the median, and the band is the central 68\% of the distribution. The dashed line is the \texttt{ROMULUS} value $A=1.17$. The horizontal band is the $M_{\star}$ range of four of the six off-nuclear TDE parents.}
    \label{fig:cdf_slope}
\end{figure}

Figure~\ref{fig:cdf_slope} shows how the host $M_{\star}$ of WBHs depends on the slope of the occupation relation, $\langle N_{\rm WBH}\rangle \propto M_{200}^{A}$. The GSMF and the SHMR are held fixed. The curve is the median $M_{\star}$ and the band is the central 68\% of the distribution. The dashed line is the \texttt{ROMULUS} slope, $A=1.17$. Over the slopes shown, that central 68\% stays at high $M_{\star}$ and continues to cover the off-nuclear parent galaxies. The horizontal band is the $M_{\star}$ range of four of the six off-nuclear parents.

\begin{figure*}[t]
\centering
\includegraphics[width=\textwidth]{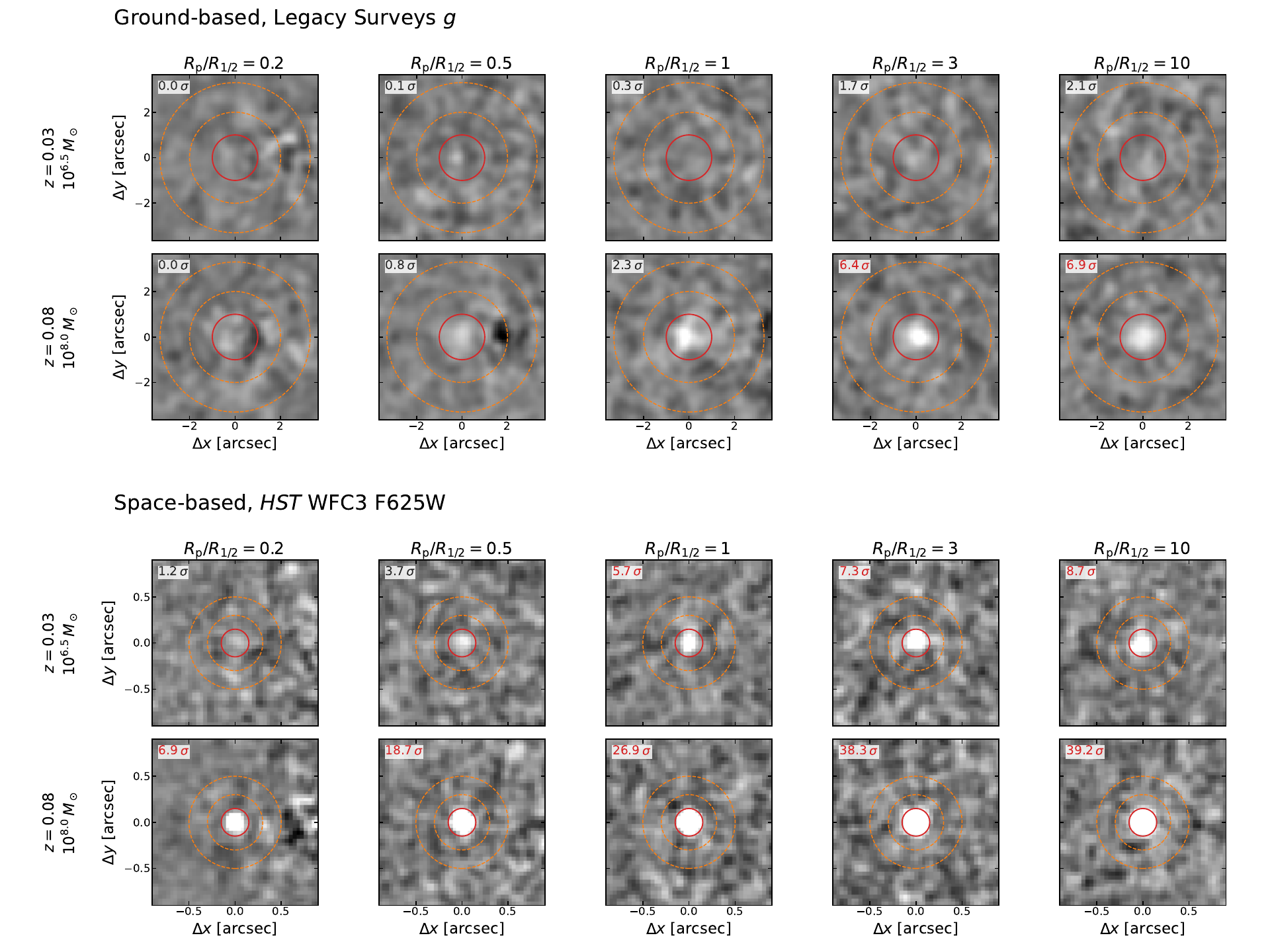}
\caption{Example residuals from the stellar-counterpart simulation of \S\ref{sec:stellar_counterparts}, after subtraction of the parent-galaxy light. The top block is Legacy Surveys $g$ and the bottom block is \hst\ WFC3 F625W. The number in each panel is the aperture signal-to-noise ratio.}
\label{fig:cp_examples}
\end{figure*}

Figure~\ref{fig:cp_examples} shows example residuals from the simulation in \S\ref{sec:stellar_counterparts}, after the parent-galaxy light has been subtracted. The top block is Legacy Surveys $g$, smoothed by half the PSF, and the bottom block is a 600~s \hst\ WFC3 F625W exposure, smoothed by the PSF. Columns run from $R_{\mathrm{p}}/R_{1/2}=0.2$ to $10$. The number in each panel is the aperture signal-to-noise ratio. The red circle marks the stellar counterpart and the orange circle the photometric aperture.

\section{Companion Counts}
\label{app:3}

The companion number $N$ used in \S\ref{sec:results_obs} is measured in the same way for every parent or host. A neighboring galaxy $i$ enters $N$ when $M_{\star,i}\ge 10^{9}\,M_\odot$, $|cz-cz_i|<5\sigma_{\rm los}$, and $R_p<R_{200}+3\delta R_{200}$. Here $cz$ and $cz_i$ are the recession velocities of the parent or host and of galaxy $i$, and $R_p$ is their projected separation. $\delta R_{200}$ is the upper $1\sigma$ uncertainty on $R_{200}$, so the search aperture is $R_{200}$ extended by three times that error. The velocity scale is the halo line-of-sight dispersion, $\sigma_{\rm los}=V_{200}/\sqrt{2}$, with $V_{200}=(GM_{200}/R_{200})^{1/2}$.

The aperture is the halo assigned to that galaxy. For an off-nuclear parent, $M_{200}$ and $R_{200}$ are the values derived in Appendix~\ref{app:1}. For a nuclear TDE host, and for a dynamical parent other than the Milky Way, $M_{200}$ is taken from the mean $z=0$ SHMR of \citet{Moster2013} at the $M_{\star}$ measured in Appendix~\ref{app:2}, and $R_{200}$ follows from the definition in Appendix~\ref{app:1}. The Milky Way uses the halo in Table~\ref{tab:dyn}.

Positions and redshifts come from the NASA/IPAC Extragalactic Database \citep{NED}. Stars, supernovae, quasars, H~\textsc{ii} regions, and objects cataloged as a group or a cluster are removed. Listings separated by only a few arcseconds are kept as one galaxy, and the spectroscopic redshift is retained. The galaxy at the center of the aperture is left out of its own count. A spectroscopic companion that passes the three cuts adds 1 to $N$.

A galaxy with only a photometric redshift contributes a weight $w$, the probability that it is a companion. These redshifts are taken from the photometric catalog of \citet{Zhou2023}. Each is represented as a Gaussian centered on the median, with widths set by the 16th and 84th percentiles. The weight is the fraction of that Gaussian for which the galaxy would lie inside the velocity window and above $10^{9}\,M_{\odot}$. The count is $N = N_{\rm spec} + \sum w$, so $N$ need not be an integer. Each weight is an independent probability, so its variance is $w(1-w)$, and $\sigma_N=\sqrt{\sum w(1-w)}$. The uncertainty is asymmetric, because $N$ cannot fall below $N_{\rm spec}$: the lower error is $\min(\sigma_N,\,N-N_{\rm spec})$ and the upper error is $\sigma_N$. A spectroscopic companion has $w=1$ and does not contribute to $\sigma_N$. The Milky Way is outside this search. Its one companion above $10^{9}\,M_{\odot}$ inside $R_{200}$ is the Large Magellanic Cloud, so $N=1$.

The $M_{\star}$ of each companion is measured from Legacy Surveys Data Release~10 photometry \citep{Dey2019} with the mass-to-light ratio of Appendix~\ref{app:2} \citep{Baldry2012}. Where the $i$ band is unavailable, $(g-i)$ and the $i$ luminosity are estimated from extinction-corrected $g$ and $r$ as in Appendix~\ref{app:2}.

The count is then used to test whether the parent or host is the most massive galaxy inside the aperture. Its mass is the $M_{\star}$ from Appendix~\ref{app:2}, and the companions use the masses just described. If a spectroscopic companion is more massive, that companion is the central galaxy and the parent or host is a satellite. The count is repeated with the aperture centered on the central galaxy, inside the $R_{200}$ given by the \citet{Moster2013} relation at the central galaxy's $M_{\star}$. These systems are the open symbols in Fig.~\ref{fig:env}.

\begin{figure}
    \centering
    \includegraphics[width=\linewidth]{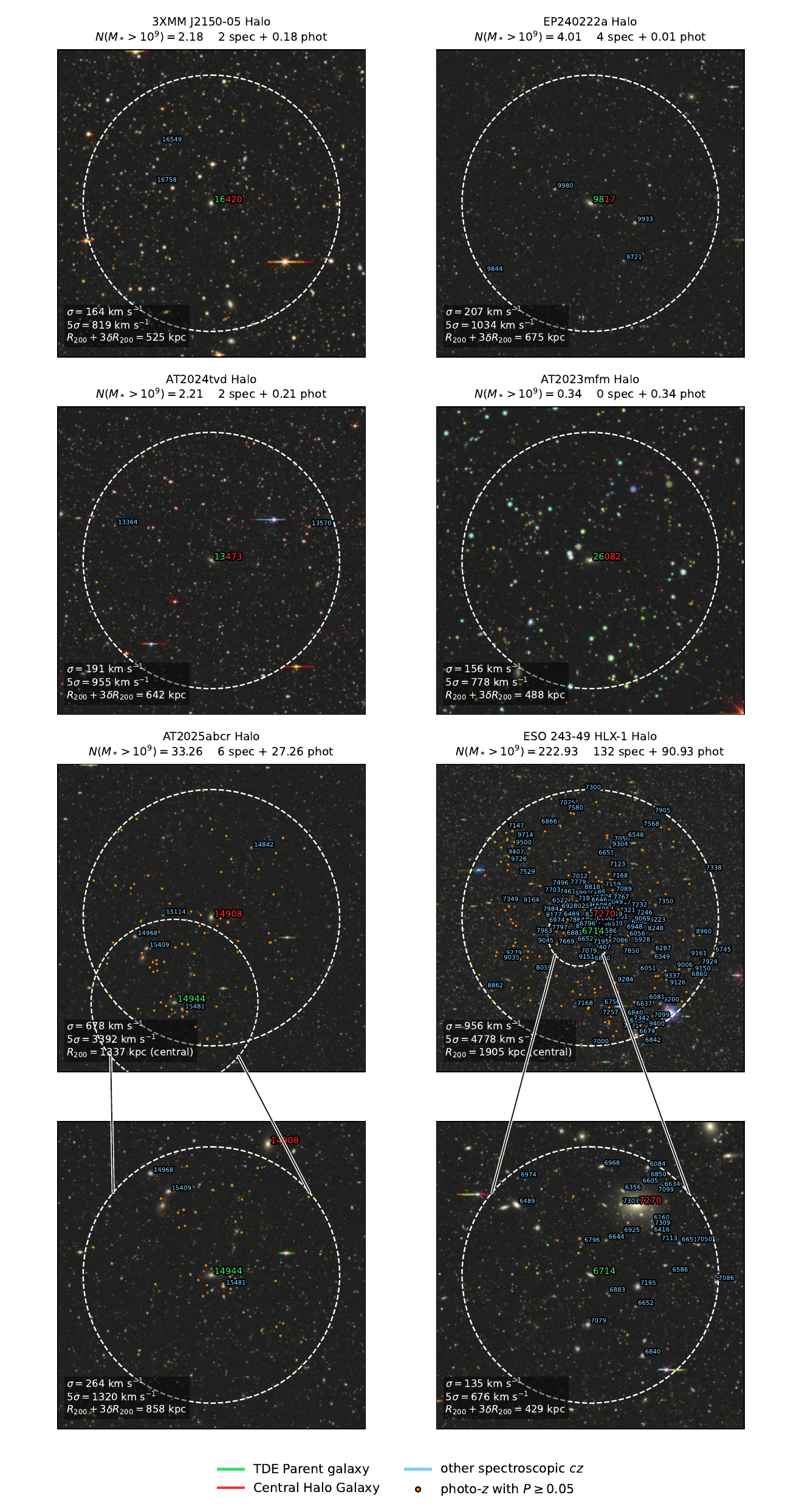}
    \caption{Legacy Surveys images of the companion search, shown for the off-nuclear parents. The dashed circle is the search aperture, and the number next to a galaxy is its recession velocity in km~s$^{-1}$. Green marks the parent, blue a spectroscopic companion, and orange a photometric-redshift candidate, drawn where the membership probability is at least 0.05. 3XMM~J2150$-$05, EP240222a, AT2024tvd, and AT2023mfm are the central galaxies of their halos. ESO~243--49 (HLX-1) and AT2025abcr are not: red marks their central galaxies, IC~1633 and MCG~$-$03-05-019. For each of those two, the upper panel is centered on the central galaxy and its $R_{200}$, the lower panel is centered on the parent, and the lines connect the parent's aperture on the wide view to the zoom. The title of each panel gives $N$ and the spectroscopic and photometric contributions.}
    \label{fig:envcut}
\end{figure}

Figure~\ref{fig:envcut} illustrates the search. The same images were made for every parent or host in the nuclear, off-nuclear, and dynamical samples. The figure shows the off-nuclear parents.

\bibliography{tde}{}
\bibliographystyle{aasjournal}

\end{document}